\documentclass[prb,aps,preprint,showpacs]{revtex4}

\usepackage{amssymb}
\usepackage{amsmath}
\usepackage{amsfonts}
\usepackage{tabularx}
\usepackage{pifont}
\usepackage{makeidx}
\usepackage[dvips]{graphicx}
\usepackage{graphics}
\usepackage{subfigure}
\usepackage{longtable}
\usepackage{dcolumn}
\usepackage{enumerate}

\begin{document}
\bibliographystyle{apsrev}

\newcommand {\bra}[1]{\langle \: #1 \: |}
\newcommand {\ket}[1]{| \: #1 \: \rangle}
\newcommand {\unity}{{\textrm{1}\hspace*{-0.55ex}\textrm{l}}}
\newcommand {\dotp}[3]{\langle \: #1 | \: #2 \: | \: #3 \: \rangle}
\newcommand {\expect}[1]{\langle \: #1 \: \rangle}
\newcommand {\ybco}[1]{YBa$_2$Cu$_3$O$_{#1}$}
\newcommand {\ybcoE}{YBa$_2$Cu$_4$O$_{8}$ }
\newcommand {\lacuo}{La$_2$CuO$_4$ }
\newcommand {\lasco}{La$_{2-x}$Sr$_{x}$CuO$_{4}$ }
\newcommand {\cuo}{CuO$_2$}
\newcommand {\Kbar}[3]{{^{#1}\overline{K}_{#2}^{#3}}}
\newcommand {\Ks}[2]{{K_{#1}^{#2}}}
\newcommand {\tone}{{^{k}\!T_{1 \alpha}^{-1}}}
\newcommand {\us}[2]{{^{#1}U_{#2}}}
\newcommand {\Vs}[2]{{^{#1}V_{#2}}}
\newcommand {\Vsn}[2]{{^{#1}V_{#2}^{0}}}
\newcommand {\Vfc}[2]{{^{#1}V_{#2}^{\mathit{fc}}}}
\newcommand {\Tm}[2]{{^{#1}T_{1#2}^{-1}}}
\newcommand {\Tone}[2]{ {^{#1}T}_{1#2}}
\newcommand {\usbar}[2]{{^{#1}\overline{U}_{#2}}}
\newcommand {\ustil}[2]{{^{#1}\tilde{U}_{#2}}}
\newcommand {\ushat}[2]{{^{#1}\hat{U}_{#2}}}
\newcommand {\TTm}[2]{\left({^{#1}T}_{1#2}T\right)^{-1}}
\newcommand {\gam}[1]{{^{#1}\gamma}}
\newcommand {\Q}[3]{{^{#1}V_{#2}^{#3}}}
\newcommand {\taus}[2]{{^{#1}\tau}_{#2}}
\newcommand {\taueff}{$\tau_{\mathit{eff}}$}
\newcommand {\taueffm}{\tau_{\mathit{eff}}}
\newcommand {\Tg}[2]{{^{#1}T_{2G#2}^{-1}}}
\newcommand {\chir}[1]{\chi_{#1}'}

\title{Analysis of NMR Spin-Lattice Relaxation Rates in 
Cuprates}

\author{A. Uldry and P. F. Meier}

\affiliation{Physics Institute, University of Zurich, CH-8057 Zurich, 
         Switzerland}

\date{\today}

\begin{abstract}
We investigate nuclear spin-lattice relaxation data in the normal state of optimally doped \ybco{7} by analyzing the contributions to the relaxation rate of the copper, planar oxygen and yttrium along the directions perpendicular to the applied field. In this new picture there is no contrasting temperature dependence of the copper and oxygen relaxation. We use the model of fluctuating fields to express the rates in terms of hyperfine interaction energies and an effective correlation time $\taueffm$ characterizing the dynamics of the spin fluid. The former contain the effects of the antiferromagnetic static spin correlations, which cause the hyperfine field constants to be added coherently at low temperature and incoherently at high temperature. The degree of coherency is therefore controlled by the spin-spin correlations. The model is used to determine the temperature-dependent correlation lengths. The temperature-dependent effective correlation time is found to be made up of a linear and a constant 
contribution that can be related to scattering and spin fluctuations of localized moments respectively. The extrapolation of our calculation at higher temperature fits the data also very well at those temperatures. The underdoped compounds \ybco{6.63} and \ybcoE are studied in the limit of the data available with some success by modifying the effective correlation time with a gap parameter. The copper data of the \lasco series are then discussed in terms of the interplay between the two contributions to $\taueffm$.
\end{abstract}
 
\pacs{74.25.-q,74.25.Ha,76.60.-k}

\maketitle



 \section{Introduction}
The nuclear resonance techniques,
nuclear magnetic resonance (NMR) and nuclear quadrupole resonance (NQR),
are powerful probes for investigating the microscopic magnetic
properties of cuprates that exhibit high-temperature 
superconductivity.
An eminent advantage of these methods is based on their highly local nature
which allows one to get information about the distinct chemical species
in the materials and make selective measurements of different crystallographic
sites (for reviews, see Refs. \onlinecite{book:mali,rev:slichter,berthier,rigamonti}). Magnetic hyperfine interactions couple the nuclear spins to the electron
system and it is essential that they are known as accurately as possible
in order to allow a correct interpretation of the properties of the electron
liquid in terms of measured NMR or NQR data. \\
One particular aspect that absorbed the attention of the NMR community is that the experiments in \ybco{7}, \ybco{6.63}, and in La$_{1.85}$Sr$_{0.15}$CuO$_4$ 
(Refs. \onlinecite{warren:87,imai1:88,imai2:88,hammel:89,warren:89,wals:90,imai:90,taki:91,wals:94})
seem to show a dramatic contrast in nuclear-spin lattice
relaxation rate behavior between copper and oxygen sites in the CuO$_2$-plane,
although these sites lie less than 2 {\AA} apart. It was concluded 
that the relaxation at the copper site exhibits strong antiferromagnetic (AFM) 
enhancement effects, whereas that at the planar oxygen site is weakly
enhanced with strikingly different temperature dependence.

The nuclear spin-lattice relaxation rate $\tone$ for a nuclear species $k$ is 
the rate at which the magnetization relaxes to its equilibrium value in the
external magnetic field applied in direction $\alpha$. The relaxation of the nuclei under consideration in the cuprates is caused by two or more fluctuating hyperfine fields that originate from magnetic moments localized near the coppers. Since the squares of these fields come into play, one of the first tasks when interpreting spin-lattice relaxation data is to determine whether the hyperfine fields should be added coherently or incoherently at the nucleus. While Mila and Rice~\cite{milarice} added them incoherently, Monien, Pines and Slichter~\cite{mps} considered both extreme cases, and from the analysis of the copper data in \ybco{7} concluded that within a one-component model the fields should be added 
coherently. The question of coherency was put aside when Millis, Monien and Pines (MMP) gave a quantitative and complete 
phenomenological description of the relevant measurements by putting
forward a model~\cite{mmp} where the nuclear spin-lattice relaxation rate, via the fluctuation-dissipation theorem, is expressed in terms of the low-frequency limit of the imaginary
part of the spin susceptibility $\chi$ 
\begin{equation}\label{moryia}
\tone \propto \sum_{q} {^k\!F_{\alpha} (\vec{q})} \lim_{\omega\to 0}
              \left( \chi^{''}(\vec{q},\omega) / \omega \right).
\end{equation}
The form factors $^k\!F_{\alpha}(\vec{q})$ depend on the geometrical arrangements of the
nuclei and the localized electronic spins. Under the form Eq. (\ref{moryia}), the question of the degree of coherency is delegated to the choice of a form for the susceptibility. 
The ``basic'' idea behind the MMP model was to account
for strong AFM correlations which exist in the cuprates 
even in the overdoped regime. The MMP model therefore postulated a spin susceptibility
which is strongly peaked at the AFM wave-vector 
$\vec{Q} = (\pi,\pi)$. In this way, the seemingly different relaxation behavior of copper and oxygen could be understood and almost all       
NMR and NQR relaxation measurements in cuprates have been analyzed using the MMP approach. \\
In a later development of the model~\cite{zha}, 
the susceptibility $\chi(q,\omega)$ was split into two parts, $\chi = \chi_{\textrm{AF}} + \chi_{\textrm{FL}}$. The first term,  $\chi_{\textrm{AF}}$ represents the anomalous  
contribution to the spin system and is peaked at or near $\vec{Q}$.
The second term, $\chi_{\textrm{FL}}$, is a parameterized form of the normal Fermi
liquid contribution. 
For copper, the contributions from the Fermi liquid part are much smaller than those from 
$\chi_{AF}$  but they dominate the relaxation of oxygen and yttrium nuclei. In the parameterization for $\chi$ introduced by MMP, $\chi_{\textrm{AF}}$ is strongly peaked at $\vec{Q}$ and the hyperfine fields at the copper are added essentially coherently. On the other hand, for any $\vec{q}$-independent
$\chi_{\textrm{FL}}$, the contributions of these fields are strictly 
incoherent, and the fields at the oxygen are therefore added incoherently in the 
MMP theory. \\

The goal of our work is to get the information on the degree of coherency directly from the data. We therefore intentionally avoid the use of Eq. (\ref{moryia}), since the degree of coherency of the hyperfine fields at a nucleus depends on details of the susceptibility which are difficult to model. Moreover, the possibility that two or more processes contribute to the spin relaxation might make it difficult to disentangle these different contributions. We therefore stay 
in the direct space since there are only a few points in the lattice
which are relevant for NMR. In this case the coherency is related to the spin-spin correlations, which we take as a parameter that we deduce from the experiments. The spin-spin correlations are found to be temperature-dependent, and therefore the degree of coherency varies with the temperature. As expected, the coherency is reduced when the temperature is increased. We note that this is also what MMP finds in the case of the copper nuclei: the peak in the susceptibility (as well as the sometimes forgotten cut-off~\cite{mmp}) gets broader as the correlations decrease. However, the approach taken here reveals other advantages: in contrast to MMP, the relaxation of both the copper and the oxygen (as well as the yttrium) is caused by the {\it same} temperature-dependent relaxation mechanism of the spin liquid, which we will characterize by an effective correlation time. Also, the different temperature behavior of the oxygen and copper nuclear relaxation rates can then be explained naturally from the particular values of the hyperfine field constants.\\

The premises of our different approach to the analysis of spin-lattice 
relaxation data in the normal state of cuprates are the following. We go back
to the simple model of fluctuating fields~\cite{book:slichter} that has been applied by 
Pennington {\it et al.}~\cite{penn} to analyze their spin-lattice relaxation rate data 
for copper in \ybco{7}. This model is particularly appropriate for anisotropic substances, as it is the case of the cuprates. We retain, however, the AFM correlations which are an
essential feature of the concept of the nearly AFM Fermi liquid. The normalized AFM spin-spin correlations between adjacent coppers are a key quantity, since they determine to what extent the hyperfine fields are added coherently.\\
We adopt the usual form for the spin Hamiltonian for copper as proposed by Mila and Rice \cite{milarice}, whereas that for the oxygen is determined by transferred contributions from the two nearest neighbor copper ions (Shastry~\cite{shastry}). We note that quantum-chemical calculations~\cite{huesser,renold} have shown
that contributions to the oxygen hyperfine interaction arising from further
distant copper ions are marginal and that an introduction of a substantial transferred field 
from next nearest neighbor Cu is not justified. \\
We will assume that all data 
obtained in the normal state of the cuprates can be attributed to purely 
magnetic relaxation although there are indications~\cite{suter} that some 
observed phenomena hint for an additional influence of charge fluctuations.\\
We confine our analysis to measurements of the planar copper and oxygen 
and the yttrium nuclei in the normal state. We neglect orthorhombicity and assume a quadratic
CuO$_2$-plane with a lattice constant of unity. For the planar oxygen we distinguish between the direction $a$ parallel to the Cu-O-Cu
bond, and $b$, perpendicular to the bond.\\

Our work is structured as follows. We introduce in Sec.~\ref{sec:dataanis} a new representation of the relaxation
data which is appropriate to their analysis in anisotropic materials. 
Sec.~\ref{sec:model} presents the model and assumptions made. We then apply our treatment to \ybco{7} in Sec.~\ref{sec:ybco7} and confront our results with a range of experiments, in particular those made at high temperature. The significance of the model parameters at low temperature are discussed. Due to lack of data, the underdoped materials \ybco{6.63} and \ybcoE are less extensively treated in Sec.~\ref{sec:underdop}. We turn to the \lasco series in Sec.~\ref{sec:lasco} and discuss the temperature and doping dependence in view of our model, in particular in the high temperature limit. We present a summary and some conclusions in Sec.~\ref{sec:sum}. Some special considerations are collected in the Appendices.

\section{Representation of data for anisotropic materials} \label{sec:dataanis}
In this section we plot the spin-lattice relaxation data in a way which suggests that the same relaxation's mechanism is at work for all the nuclei under consideration.\\
The magnetic relaxation process into equilibrium is caused by 
fluctuating effective magnetic fields along the two
orthogonal axes $\beta$ and $\gamma$
perpendicular to the direction $\alpha$ of the applied field, which we write as
\begin{equation}\label{defTm}
  \Tm{k}{\alpha}={\us{k}{\beta}}+{\us{k}{\gamma}}
\end{equation}
where $\left( \alpha, \beta, \gamma \right) = (x,y,z)$. The quantities $\us{k}{\alpha}$ describe then the contribution to $\Tm{k}{\beta}$ and $\Tm{k}{\gamma}$ caused by fluctuating fields in the crystallographic direction $\alpha$.
From (\ref{defTm}) the $\us{k}{\alpha}$ are therefore given by
\begin{equation}\label{defU}
{\us{k}{\alpha}}= \frac{1}{2}
\left[ -{\Tm{k}{\alpha}}+{\Tm{k}{\beta}}+{\Tm{k}{\gamma}}\right] 
\end{equation}
These transformed rates $\us{k}{\alpha}$, which in the following will just be called rates, are not directly accessible by experiment
except for particular symmetries (e.g., $\us{63}{ab} = \Tm{63}{c}/2$),
but in general can be obtained if a complete set of data measured with the 
applied field along all three crystallographic axes is available.
At variance with the 
common use to represent NMR spin-lattice relaxation rates in  $(T_1T)^{-1}$
versus $T$ plots, we prefer here to study the rates $\us{}{\alpha}$. 
This change
in representation is trivial. It was however pointed out 
previously by one of us~\cite{hono} that it is much more instructive 
to investigate
ratios between $\us{k}{\alpha}$ for $\alpha$ in different directions than 
ratios of the corresponding rates $\Tm{k}{\alpha}$.

The practice to represent and analyze $(T_1T)^{-1}$ data grew out from 
applications in liquids where the rates are isotropic and $(T_1T)^{-1}$
is temperature independent in simple metals. For layered cuprates, however,
which are anisotropic materials, the analysis of NMR data in terms of 
${\us{k}{\alpha}}$ has distinct advantages over that in terms of  
${\Tm{k}{\alpha}}$.

For cuprates, the most complete set of NMR and NQR data with respect to different 
nuclei and directions of external field relative to the crystallographic axes
 is available for optimally doped \ybco{7} in the normal state between 100 K and 
room temperature. We start, therefore, our analysis in this temperature range
and use the copper data for $\Tm{63}{c}$
from Hammel {\it et al.}~\cite{hammel:89}  and $\Tm{63}{ab}$ from 
Walstedt {\it et al.}~\cite{wals:88}. The oxygen data $\Tm{17}{\alpha}$
were taken from Martindale {\it et al.}~\cite{martind:98} 
and the yttrium data
 $\Tm{89}{\alpha}$ from Takigawa {\it et al.}~\cite{taki:93}.

\begin{figure}
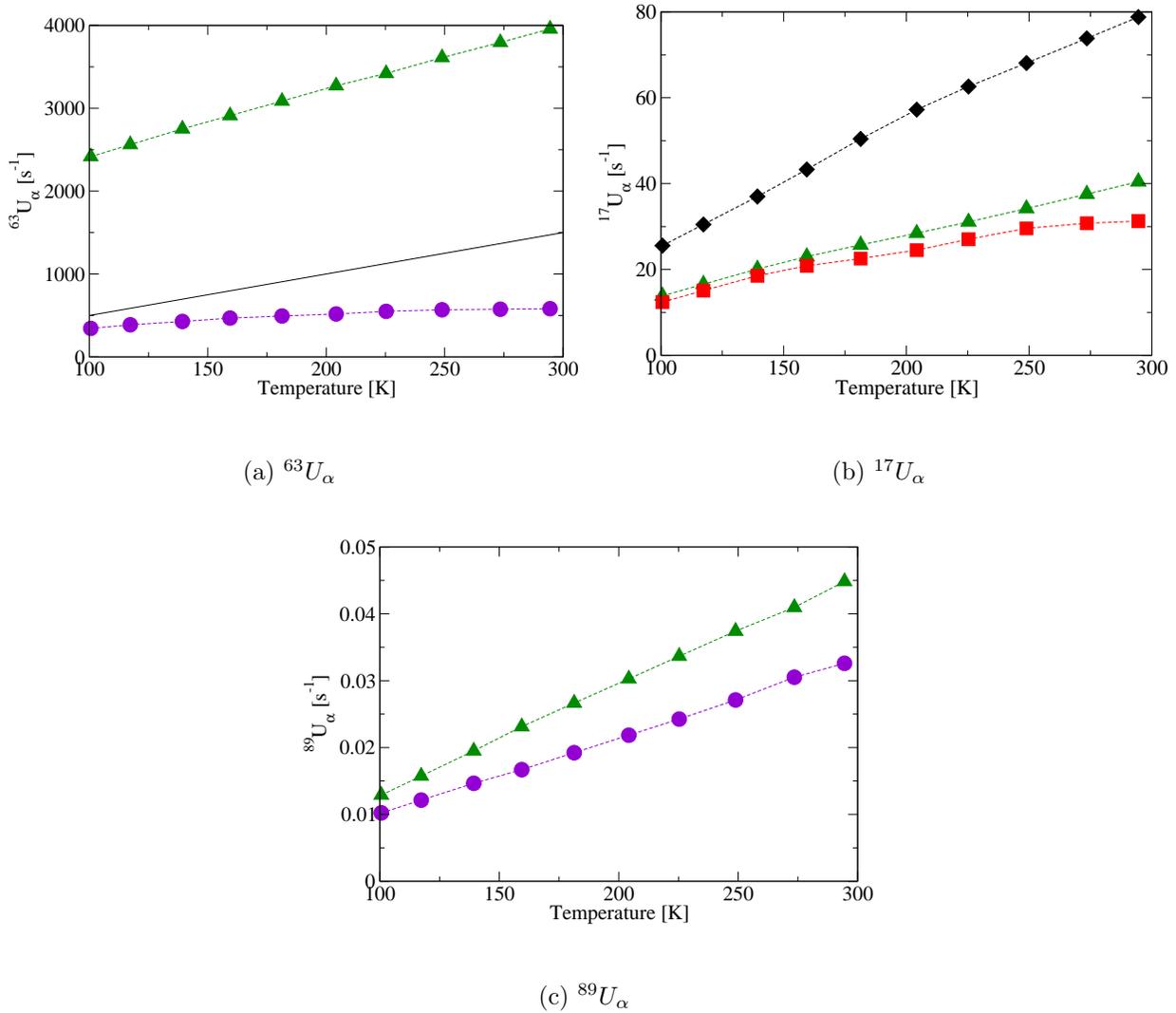

\begin{center}
\subfigure[$^{63}U_{\alpha}$]{\label{fig:CuUs}
\includegraphics*[width=8cm]{Us63_Y7.eps}}
\subfigure[$^{17}U_{\alpha}$]{\label{fig:OUs}
\includegraphics*[width=8cm]{Us17_Y7.eps}}
\subfigure[$^{89}U_{\alpha}$]{\label{fig:YUs}
\includegraphics*[width=8cm]{Us89_Y7.eps}}
\caption{Relaxation rates $\us{k}{\alpha}$ from measurements in \ybco{7}. 
The symbols denote interpolations of $\Tm{}{\alpha}$ data~\cite{hammel:89,wals:88,martind:98,taki:93} transformed according to (\ref{defU}), in the direction $a$ (diamonds), 
$b$ (squares), $ab$ (circles), $c$ (triangles).
Dashed lines are guides for the eyes. The full straight line in (a)
denotes the values for metallic copper.
\label{fig:Us}}
\end{center}
\end{figure}

These sets of data for the rates $\Tm{k}{\alpha}$, interpolated at the same temperature points, allow the transformation into our rates $\us{k}{\alpha}$ according to (\ref{defU}). Neglecting any 
error analysis, the results for the three nuclei are shown 
in Fig.~\ref{fig:Us}. 

In these representations of the data there is nothing to see of a drastic
contrast in the
temperature dependence of the relaxation rates of copper and oxygen
nuclei in the same CuO$_2$ plane which has, as mentioned in the Introduction,
intrigued the NMR community.
All relaxation rates grow with increasing temperature as is expected for fluctuations.
The differences in the magnitude for Cu, O, and Y are due to different 
strengths of the hyperfine interaction energies which allow a scaling of all 
data as is shown in Appendix~\ref{app:Ubars}.
We stress that so far no model is used, and nothing else was done other than adding and subtracting the data.

The relaxation rate data, transformed now into a representation appropriate for
layered structures, suggest therefore that all nuclei under consideration relax in a {\it similar} fashion by the same mechanism of the spin liquid and that the strength of this
relaxation initially increases linearly with temperature (note that the additional line drawn 
in Fig.~\ref{fig:CuUs} depicts the relaxation measured in copper metal). The question of why the relaxation rates between the copper and the oxygen are nonetheless {\it different} will be investigated 
in \ref{sub:V}. In the following section we outline a robust and unbiased model that will allow us to 
deduce more details about the origin and the source of the temperature dependence
of the spin-lattice relaxation. 

\section{Model and assumptions}\label{sec:model}
The quantities $\us{k}{\alpha}$ have been 
introduced as the contributions from
fluctuating local effective magnetic fields $H_{\alpha}$ along the direction
$\alpha$. In this section we adopt the calculation of relaxation rates in terms of fluctuating fields described in Ref. \onlinecite{book:slichter} to determine an expression for $\us{k}{\alpha}$ in the present case.
In particular in the cuprates the hyperfine interaction energies depend on the static
AFM spin correlations. We find that $\us{k}{\alpha}$, for any nuclei $k$ under consideration, can be written as a static term containing the spin correlations and hyperfine field constants, and another which is the effective correlation time.\\

\begin{figure}
\begin{center}
\includegraphics*[width=4cm]{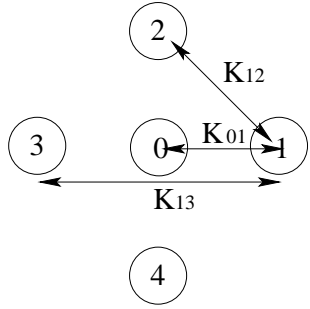}
\caption{Schematic picture of the AFM spin correlations $\Ks{ij}{}$ between electron moments at the copper sites $0$, $1$,.. $4$. \label{fig:korrel}}
\end{center}
\end{figure}

Let us consider first an oxygen nucleus with spin $\,{^{17}I}\,$ and gyromagnetic ratio $\gam{17}$, for which the hyperfine interaction Hamiltonian is determined by
\begin{equation}
{^{17}\mathcal{H}}(t)=-\hbar \,\,\gam{17} \, \,{^{17}\vec{H}}(t)\,\cdot \,^{17}\vec{I}\, .
\end{equation}
The field operator $\,{^{17}\vec{H}}(t)\,$ for an oxygen situated between site $0$ and $1$ (Fig. \ref{fig:korrel}) is assumed to originate from magnetic moments with spin $S=1/2$ localized on those two adjacent nearest neighbor (NN) copper ions with spin components $S^{0}_{\alpha}$ and $S^{1}_{\alpha}$ 
respectively (Fig. \ref{fig:korrel}). 
The field operator components are given by
\begin{equation}
{^{17}H}_{\alpha}(t)=-\hbar\, \gamma_e \, c_{\alpha}[S^{0}_{\alpha}(t)+S^{1}_{\alpha}(t) ]
\end{equation}
where $c_{\alpha}$ is the diagonal element of the hyperfine tensor in
direction $\alpha$ given in units of spin densities ($a_B^{-3}$).\\
Provided that the fluctuations in different field directions are independent, the components of the auto-correlation function of $\,{^{17}H}(t)\,$ 
are 
\begin{equation}
\langle ^{17}H_{\alpha}(t)\,^{17}H_{\alpha}(0) \rangle
 =  \hbar^2\,\gamma^{2}_{e}\,c^{2}_{\alpha}\: 2\: \langle S_0^{\alpha}(t) S_0^{\alpha}(0)+ S^{\alpha}_0 (t) S^{\alpha}_1 (0)\rangle .
\end{equation}
Assuming an exponential decay of correlations in time, the expression in brackets can be written as
\begin{equation}\label{deftaueff}
  \langle S_0^{\alpha}(t) S_0^{\alpha}(0) + 
     S^{\alpha}_0 (t) S^{\alpha}_1 (0) \rangle = \frac{1}{4}
        \left( 1 + K^{\alpha}_{01} \right)e^{-|t|/\tau_{\mathit{eff}}}. 
\end{equation}
$\tau_{\mathit{eff}}$ is an effective correlation time that acts as a time-scale for the fluctuations~\cite{book:slichter}. Its temperature dependence will give us an indication as to what type of processes come into play in the relaxation. $K^{\alpha}_{01}$ is the normalized
``static'' NN spin correlation
\begin{equation}
   K^{\alpha}_{01} = 4 \left< S_0^{\alpha}(0) S_1^{\alpha}(0) \right> .
\end{equation}
The correlations of two NN moments is thus separated into a spatial AFM correlation $K^{\alpha}_{01}$ that will determine the degree of coherency and temporal correlation that characterizes the dynamics of the electronic spin fluid system which exchanges energy with the nuclei.\\
We make the assumption that the correlation time $\tau_{\mathit{eff}}$ is isotropic and in the following that it is also independent
of the spatial separation of the correlated spins. However 
$K^{\alpha}_{01}$ may be different for in-plane and out-of-plane components.
Following Slichter~\cite{book:slichter}, we find that $\us{k}{\alpha}$ is obtained as
\begin{equation}
\us{k}{\alpha}=\frac{1}{2\hbar^2}\int_{-\infty}^{\infty}\langle ^{k}H_{\alpha}(t)\,^{k}H_{\alpha}(0) \rangle e^{i\omega_kt}\mathit{dt}
\end{equation}
where $\omega_k$ is the Larmor frequency.
The expression for the contribution $ ^{17}U_{\alpha} $ to the relaxation
rate then is given by
\begin{equation}\label{uuuu}
  \us{17}{\alpha} = \frac{2 C^2_{\alpha}}{4 \hbar^2} 
\left( 1 + K^{\alpha}_{
01} \right) \tau_{\mathit{eff}}
\end{equation}
when $|\omega_k\taueffm|\ll1$, with $  C_{\alpha} = \hbar\, { \gam{17}} \,\hbar \, \gamma_e \, c_{\alpha}$.

The hyperfine field at the planar copper nucleus is determined by an on-site
contribution $A_{\alpha}$ from the copper ion with spin component $S^{\alpha}_0$
and transferred contributions $B_{\alpha}$ from the four NN ions 
with spin components $S^{\alpha}_j$, where $j = 1, 2, 3, 4$ (Fig. \ref{fig:korrel}). 
We note that ab-initio calculations of the hyperfine interactions~\cite{renold,huesser}
yield besides the isotropic transferred field $B$ also a transferred dipolar
field $B_{dip,\alpha}$ which, for simplicity, will be ignored in the following.
The corresponding equation for $^{63}U_{\alpha}$ then reads
\begin{widetext}
\begin{equation}\label{defUCu}
  \us{63}{\alpha}= \;\frac{1}{4 \hbar^2}\; \left( A^2_{\alpha} + 
4 B^2 + 8 A_{\alpha} B K^{\alpha}_{01} 
 + 8  B^2 K^{\alpha}_{12} + 4 B^2 K^{\alpha}_{13} \right) \tau_{\mathit{eff}} 
\end{equation}
\end{widetext}
where $K^{\alpha}_{12}$ and $K^{\alpha}_{13}$ are the normalized spin 
correlations between two copper ions which are $\sqrt 2$ and 2 lattice units
apart respectively (Fig. \ref{fig:korrel}). 

In the YBaCuO compounds, the yttrium is located between two 
adjacent CuO$_2$ planes
and has four copper NN in each. Here, the situation is not as clear as for 
the Cu and O hyperfine interactions since there may be interplane spin 
correlations between copper moments and a direct dipolar coupling of the 
same order magnitude as the transferred fields. We ignore these complications 
and use the simplest form which leads to
\begin{equation}\label{defUY}
  \us{89}{\alpha} = \frac{8\; D^2_{\alpha}}{4 \; \hbar^{2}}
 \left( 1 + 2 K^{\alpha}_{01} + K^{\alpha}_{12} 
 \right) \tau_{\mathit{eff}} .
\end{equation}

It is convenient to express the relaxation rates as
\begin{equation}\label{defUO}
  \us{k}{\alpha} (T) = \,^{k}V_{\alpha}(T)\, \tau_{\mathit{eff}} (T) 
\end{equation}
with
\begin{widetext}
\begin{subequations}\label{Vsdef}
\begin{eqnarray}
  \Vs{17}{\alpha}(T) &=& \;\frac{1}{4 \hbar^2}\; 2 C^2_{\alpha}
\lbrack 1 + K^{\alpha}_{01}(T)  \rbrack\\
  ^{63}V_{\alpha}(T) &=& \;\frac{1}{4 \hbar^2}\; \lbrack A^2_{\alpha} + 
4 B^2 + 8 A_{\alpha} B K^{\alpha}_{01} (T)
 + 8  B^2 K^{\alpha}_{12}(T) + 4 B^2 K^{\alpha}_{13}(T) \rbrack \label{VsdefCu}\\
  \Vs{89}{\alpha}(T) &=&  \;\frac{1}{4 \hbar^2}\; 8 D^2_{\alpha}
 \lbrack 1 + 2 K^{\alpha}_{01}(T) + K^{\alpha}_{12}(T) 
 \rbrack .
\end{eqnarray}
\end{subequations}
\end{widetext}
This factorization emphasizes the different temperature dependencies that determine
$ ^{k}U_{\alpha} (T)$.
The $^{k}V_{\alpha} (T)$, which apart from the factor $\hbar^2$ are the static
hyperfine energies squared, vary with temperature due to changes of the static
spin correlations $K^{\alpha}_{ij}(T)$, whereas 
the effective correlation time $\taueffm(T)$ reflects the 
changes in the dynamics. 

We denote the limiting values when all correlations are zero by $\Vsn{k}{\alpha}$ 
\begin{subequations}\label{V0}
\begin{eqnarray}
\Vsn{63}{\alpha} &=& \frac{1}{4 \hbar^2} \left(A_{\alpha}^2 
+ 4 B^2 \right)\label{V063}\\
\Vsn{17}{\alpha} &=& \frac{1}{4 \hbar^2} 2 C_{\alpha}^2  \label{V017} \\
\Vsn{89}{\alpha} &=& \frac{1}{4 \hbar^2}8D^{2}_{\alpha} \label{V089}
\end{eqnarray}
\end{subequations}
and by  $\Vfc{k}{\alpha}$ for full AFM correlations:
\begin{subequations}\label{Vfc}
\begin{eqnarray}
\Vfc{63}{\alpha}  &=& \frac{1}{4 \hbar^2} 
\left(A_{\alpha} - 4 B \right) ^2 \label{Vfc63}\\
\Vfc{17}{\alpha} &=& 0 \label{Vfc17}\\
\Vfc{89}{\alpha} &=& 0\label{Vfc89}.
\end{eqnarray}
\end{subequations}
$\Vs{k}{\alpha}(T)$ accounts for the transition from the fully correlated situation, where the hyperfine fields are added coherently, towards the completely uncorrelated regime given by $\Vsn{k}{\alpha}$, where the fields are added incoherently.\\

We would like to point out that the model we use is an extension of a well established approach. It has been applied, e.g. by Monien, Pines and Slichter~\cite{mps} to analyze $\Tm{63}{}$ data in the limiting cases $\Vsn{63}{}$ and $\Vfc{63}{}$. The new feature is to adopt the existence of AFM correlations in the cuprates that are static with respect to typical NMR times and vary with the temperature, and this also has been done before, as it is detailed in section \ref{sub:spinspin}.\\

In order to reduce the number of parameters we assume that the static
spin correlations are antiferromagnetic and that  $K^{\alpha}_{12}$
and $K^{\alpha}_{13}$  depend directly on the value of $K^{\alpha}_{01}$, according to
\begin{equation}\label{Ks}
K^{\alpha}_{12} = \mid K_{01}^{\alpha} \mid^{\sqrt2} \quad \textrm{and}\quad
     K^{\alpha}_{13} = \mid K_{01}^{\alpha} \mid ^2.
\end{equation}
This particular choice of an exponential decay with length $\lambda_{\alpha}$, defined so that 
\begin{equation}
     K^{\alpha}_{01} = - \exp \left[ - 1/ \lambda_{\alpha} \right], 
\end{equation}
is guided by results of solutions of the planar anisotropic
Heisenberg model which were obtained by direct diagonalization of the
Hamiltonian for small systems~\cite{hoechner} and which suggested 
$K_{12}^{\alpha} \approx \mid K_{01}^{\alpha} \mid ^{1.5}$ and anisotropic
correlations.

These assumptions certainly influence the actual values for 
$\lambda_{\alpha}$ which will result from the analysis of the data, but 
they provide a reasonable starting point for the analysis and may be easily modified.
\section{Results for \ybco{7}}\label{sec:ybco7}
We present in this section the analysis of the optimally doped \ybco{7} data in the framework of our model. We first outline the fitting procedure, which allows us to determine the correlation lengths $\lambda_{ab}$ and $\lambda_{c}$, and the hyperfine field constants. We can then determine the effective correlation time $\taueffm$ and parameterize it in order to identify the underlying mechanisms of the spin fluid. We discuss in particular the low temperature limit, which exhibits a Fermi-liquid character for all three nuclei (Cu, O and Y). The model predictions at high temperature are then compared with experiments. Then, by looking at the extremal values of $\Vs{k}{\alpha}$, we will discuss why the measured relaxation rates of the copper and oxygen have different temperature dependence. Finally, we apply our model to spin-spin relaxation measurements.

\subsection{Fitting procedures}\label{ssec:fitting}
We define the following independent ratios $r_j$     
\begin{widetext}
\begin{eqnarray}
r_1=\frac{^{63}U_{c}}{^{17}U_{c}} \quad r_2=\frac{^{63}U_{ab}}{^{17}U_{a}}
\quad  r_3=\frac{^{63}U_{ab}}{^{17}U_{b}}
\quad  r_4=\frac{^{17}U_{b}}{^{17}U_{c}}
\quad  r_5=\frac{^{89}U_{c}}{^{89}U_{ab}}
\quad  r_6=\frac{^{17}U_{c}}{^{89}U_{c}}
\end{eqnarray}
\end{widetext}
and denote with $r_j^{\mathit{exp}}$ the ratios obtained from the experimental values
for $\us{k}{\alpha}$ as calculated from (\ref{defU}) and plotted in Fig. \ref{fig:Us}. We also form
the ratios $r_j^{\mathit{mod}}$ according to the model (Eqs. \ref{uuuu},\ref{defUCu},\ref{defUY}), for which the 
correlation times $\tau_{\mathit{eff}}$ cancel ($r_1^{\mathit{mod}}=\Vs{63}{c}/\Vs{17}{c}$ etc).

\begin{figure}
\begin{center}
\includegraphics*[width=8cm]{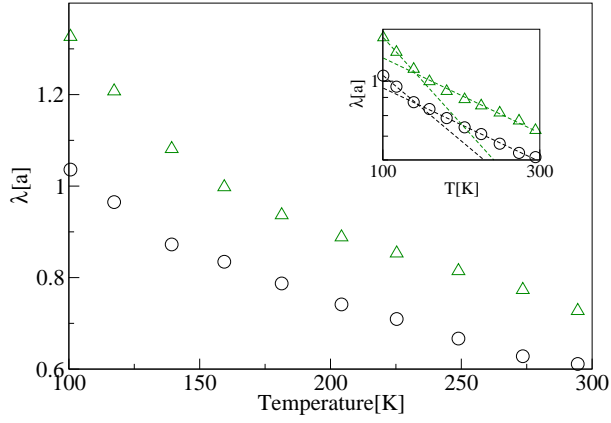}
\caption{Correlation lengths $\lambda_{\alpha}$ obtained from the fit in 
$\mathrm{Y} \mathrm{Ba}_2 \mathrm{Cu}_3 
\mathrm{O}_{7}$, in the direction ab (circles) and c (triangles). Inset: the same on a logarithmic scale, fitted with exponential functions. \label{fig:TLam4}}
\end{center}
\end{figure}

\begin{figure}
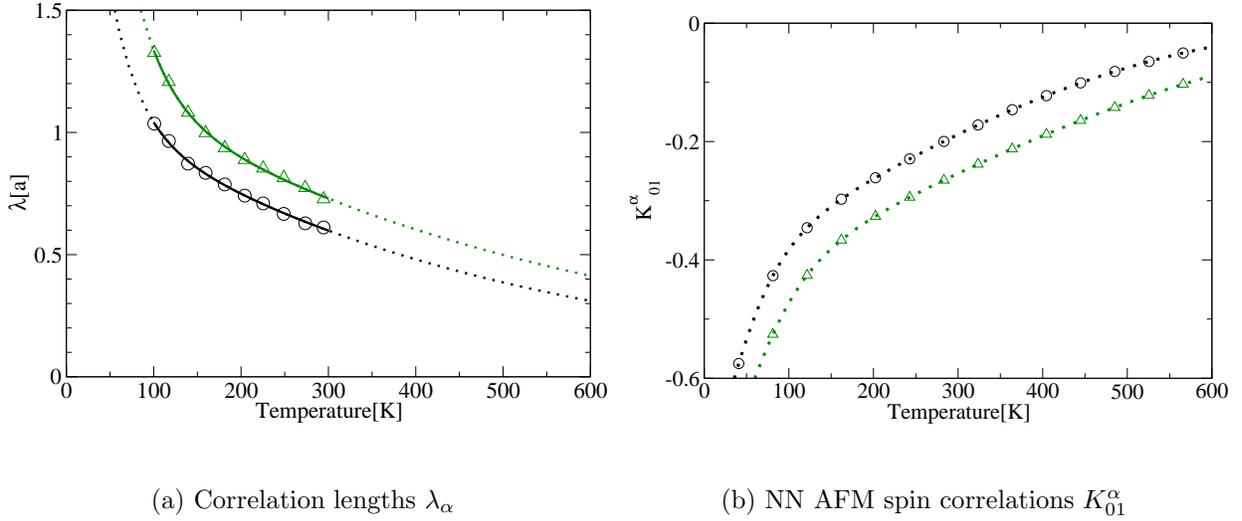

\begin{center}
\subfigure[Correlation lengths $\lambda_{\alpha}$]{\label{fig:Tlamb4fit}
\includegraphics*[width=8cm]{Tlamb4extrap.eps}}
\subfigure[NN AFM spin correlations 
$K_{01}^{\alpha}$]{\label{fig:Tcorrel4}
\includegraphics*[width=8cm]{K01Y7.eps}}
\caption{Correlations and correlation lengths from the fit (lines) in the direction ab (circles) and c(triangles). The dotted lines are extrapolations with exponential functions. \label{fig:correlT4}}
\end{center}
\end{figure}

We define the following function to minimize:
\begin{equation}\label{chisq}
\chi^2=\frac{1}{n_r}\sum_{i}^{n_r}\sum_{j}^{n_p}\left( \frac{r_i^{\mathit{exp}}(T_j)
-r_i^{\mathit{mod}}(T_j)}{r_i^{\mathit{exp}}(T_j)} \right)^2
\end{equation}
where $n_r$ (6 in the case of \ybco{7}) is the number of ratios available 
and $j$ runs through the temperature points. The function measures 
the normalized difference between the calculated ratios $r_i^{\mathit{mod}}$ and 
the experimental points $r_i^{\mathit{exp}}$ at each temperature.
We then minimize (\ref{chisq}) in order to obtain the best local solution 
for the whole set of parameters (hyperfine interaction energies and values
for the correlation lengths).
As input parameters for the hyperfine interaction energies we used the
values determined by H\"ochner~\cite{hoechner}.
The $\lambda_{ab}$ and $\lambda_c$ obtained for each temperature point 
are illustrated in Fig.~\ref{fig:TLam4} and the resulting values for 
the hyperfine interaction energies
\footnote{One can identify three contributions to the components of the general hyperfine interaction tensor. They are the isotropic (Fermi contact) term, the traceless dipolar term and the spin-orbit coupling term. We note that the core polarization is not an observable. We use instead the total Fermi contact interaction which 
can consist of on-site as well as transfered contributions. The mechanisms of spin transfer and the radial dependence of the difference between spin-up and spin-down densities at the oxygen and the copper have been illustrated by clusters calculations~\cite{renold}.}
are given in Table \ref{tab:hfeV}.
It is seen in Fig. \ref{fig:TLam4} that $\lambda_c$ is consistently somewhat 
larger than $\lambda_{ab}$ ($\lambda_c \approx 1.2 \lambda_{ab}$)
which could be related to our neglect of anisotropies in the
g-factors. From the logarithmic 
representation in the inset of Fig.~\ref{fig:TLam4} we see that the temperature dependence obtained for the $\lambda_\alpha$ can be parameterized by a 
sum of two exponential functions. This parameterization has been only 
used to extrapolate at higher and lower temperatures (dotted lines 
Fig. \ref{fig:Tlamb4fit}). 
We would like to emphasize that we are not at this stage looking into interpreting the temperature dependence of $\lambda_{\alpha}$. We merely wish the $\lambda_{\alpha}$ to be optimally fitted (regardless of the significance of the parameters), and this particularly at high temperature.

The NN
spin correlation functions $K_{01}^{\alpha}$ are plotted in 
Fig.~\ref{fig:Tcorrel4}. The correlation $\Ks{01}{ab}$ is about $-0.4$ at $T$= 100 K and drops to $-0.2$
at room temperature. The extrapolation to higher temperatures shows that 
even at $T$ = 600 K small AFM correlations ($\approx -0.04$)
exist. It is clear that these specific values depend on the particular choice (Eq.~\ref{Ks}) of exponential dependence of the correlations with the distance.

We note that the hyperfine interaction energies in Table \ref{tab:hfeV} have been only determined
from relaxation data without recourse to static measurements except
that some reasonable initial guess had to be assumed. 
It is astonishing that the resulting values are very close to those 
which have been compiled by Nandor {\it et al.}~\cite{nandor:99} and which 
are included in Table \ref{tab:hfeV}.
\begin{table*}
\begin{ruledtabular}
\begin{tabular}{cccccccccc}
  & $^{63}A_c$ & $^{63}A_{ab}$ & $^{63}B$ & $^{63}B_{\mathit{dip},c}$ & $^{17}C_a$ & $^{17}C_b$ & $^{17}C_c$ & $^{89}D_{ab}$ & $^{89}D_c$ \\
\hline
Fit & $-1.68$ & $0.168$ & $0.438$ & - & $0.259$ & $0.173$ & $0.19$6 & 
$-0.00280$ & $-0.00349$ \\
Ref.~\cite{nandor:99} & $-1.$6 & $0.29$ & $0.4$ & - & $0.25$ & $0.1$3 & $0.156$ & $-0.0048$ & $-0.0048$\\
\footnote{Cluster calculations. The error intervals are due to uncertainties in the spin-orbit interaction.}Ref.~\cite{renold} & $-1.72 \pm 0.42$ & $0.21 \pm 0.11$ & $0.29$  & $0.05$  & $0.289$  & $0.140$  & $0.151$  & -  & - \\
\end{tabular}
\end{ruledtabular}
\caption{Magnetic hyperfine interaction energies in units 
of $10^{-6}$ eV \label{tab:hfeV} }
\end{table*}

For further reference we collect in Table \ref{tab:V0} the values obtained for $\Vsn{k}{\alpha}$ calculated from (\ref{V0}) using the hyperfine interaction energies given in Table \ref{tab:hfeV}.
\begin{table}
\begin{ruledtabular}
\begin{tabular}{ccccccc}
$\Vsn{17}{a}$ & $\Vsn{17}{b}$ & $\Vsn{17}{c}$ & $\Vsn{63}{ab}$ & $\Vsn{63}{c}$ & $\Vsn{89}{ab}$ & $\Vsn{89}{c}$ \\
$77.7 $ & $34.6$ & $44.2$ & $459$ & $2080$ &$0.0361$ &$0.0562$ \\
\end{tabular}
\caption{$\Vsn{k}{\alpha}$ in units of $10^{15} s^{-2}$ \label{tab:V0}}
\end{ruledtabular}
\end{table}

\subsection{Consistency check}\label{sub:consistency}
\begin{figure}
\begin{center}
\includegraphics*[width=8cm]{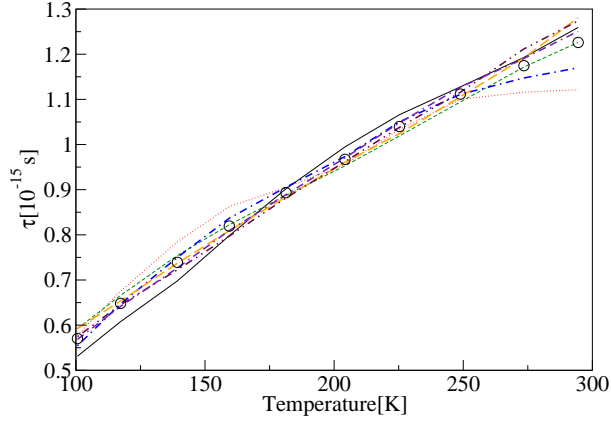}
\caption{Relaxation times obtained from the experimental rates 
$\us{k}{\alpha}$ using the fitted correlations and hyperfine fields.
The circles are the calculated average. $\taus{17}{a}$ (thin full line), 
 $\taus{17}{b}$ (thin dotted), 
 $\taus{17}{c}$ (thin dashed), 
 $\taus{63}{ab}$ (thick dash-dotted), 
 $\taus{63}{c}$ (thick dashed), 
 $\taus{89}{ab}$ (thick dash-dot-dot-dashed), 
 $\taus{89}{c}$ (thick dot-dash-dash-dotted). \label{fig:tauT4det}}
\end{center}
\end{figure}

In the previous subsection we have formed the ratios $r_i^{mod}$ in order 
to get rid of the effective correlation time $\tau_{\mathit{eff}}$ in the 
model. Having now the fitted values for the hyperfine interaction parameters 
and for $K_{01}^{\alpha}(T)$ and thus calculated the values $\Vs{k}{\alpha}$ (Eqs. \ref{Vsdef}), we get back to the experimental spin-lattice relaxation rates $\us{k}{\alpha}$. From (\ref{defUO}) we can obtain the 
resulting seven experimental values 
$^k\tau_{\mathit{eff},\alpha}$, which, according to the model assumptions,
should all be equal. These seven values for $\tau_{\mathit{eff}}$ 
are plotted in
Fig.~\ref{fig:tauT4det} in function of the temperature.

The consistency is surprisingly good. We therefore calculated the mean values (circles in Fig. \ref{fig:tauT4det}, neglecting
any weighting according to the estimated precision of the data
for the various nuclei).\\
The temperature dependence of the average $\tau_{\mathit{eff}}$ is well fit 
by a function of the form
\begin{subequations}\label{tau1nogap}
\begin{gather}
\tau_{\mathit{eff}}^{-1} = \tau_1^{-1} + \tau_2^{-1}\label{taumod}\\
\textrm{with} \quad \tau_1 = a T  \quad \textrm{and}\quad  \tau_{2} = \textrm{constant.}
\end{gather}
\end{subequations}
We obtain $a=7.0 \times 10^{-18}$ s/K
and the temperature independent $\tau_2 = 3.0 \times 10^{-15}$ s.
\begin{figure}
\begin{center}
\includegraphics*[width=8cm]{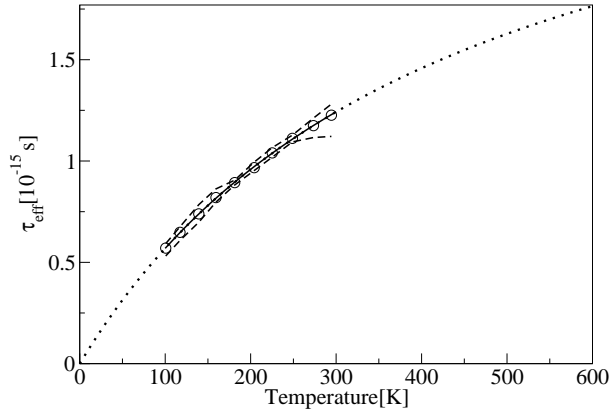}
\caption{ Temperature dependence of the averaged relaxation time $\tau_{\mathit{eff}}$.
The circles are the calculated average, and the dashed lines represent 
the upper and lower bound. The dotted line is the fit as explained in 
the text. \label{fig:tauT4}}
\end{center}
\end{figure}
The result of the fit is shown 
in Fig.~\ref{fig:tauT4} together with extrapolations 
to lower and higher temperatures.

\subsection{The ``basic'' relaxation mechanism}\label{sub:basicU}
We would like to point out that our analysis reveals something quite different from any previous analysis. The fact that all the $\taus{k}{a}$ {\it effectively} fall onto the same line demonstrates that the relaxation of all the three nuclei under consideration is governed by the same mechanism. 

In our picture, there exists what we will call a ``basic'' electronic relaxation mechanism, caused by the spin fluid, that affects the localized moments and exchanges energy with the nuclei. This mechanism is characterized by the short effective correlation time $\taueffm$. In addition, on a long time scale, the spin correlations between these moments vary with the temperature. The observed nuclear spin-lattice relaxation rate $\us{k}{\alpha}$ reflects the temperature dependence of both the ``basic'' electronic relaxation and the change in spin correlations.\\
In our model, we can disentangle these two contributions by introducing a ``basic'' nuclear relaxation rate $\ushat{k}{\alpha}(T)$, defined by
\begin{equation}
\ushat{k}{\alpha}(T)=\frac{\Vsn{k}{\alpha}}{\Vs{k}{\alpha}(T)}\,\us{k}{\alpha}(T)=\Vsn{k}{\alpha}\,\taueffm(T).
\end{equation}
The temperature dependence of the ``basic'' nuclear rate $\ushat{k}{\alpha}(T)$ is therefore that of the ``basic'' electronic relaxation mechanism by definition. Important information on the electronic system is of course also drawn from the temperature dependence of the spin correlation $\Ks{01}{\alpha}(T)$ (and hence of $\Vs{k}{\alpha}(T)$) and will be discussed in \ref{sub:V}.\\
From the consistency check (\ref{sub:consistency}), we therefore conclude that the temperature dependence of the ``basic'' relaxation rate 
\begin{enumerate}[(i)]
\item
is the same for Cu, O and Y  
\item
is linear in $T$ at low temperature.
\end{enumerate}
\subsection{Two ``basic'' electronic relaxation mechanisms}\label{sub:taueff}
There are four points which would now require a discussion. The correlation time $\tau_1$ that dominates the rates $\ushat{k}{\alpha}$ at low temperatures, the correlation time $\tau_2$ that dominates at high temperature, the crossover between the two regimes, and the particular form of $\taueffm$ (Eq. \ref{taumod}), where the rates are combined. There are various possible explanations of the crossover from the initially linear behavior to a saturation at high temperature. We confine ourselves to some ``basic'' and simple arguments in order to avoid an over-interpretation at this stage.\\

We address first the low temperature regime and consider only the contribution $\tau_1=aT$. This corresponds to the behavior of $\Tm{}{}$ in a metallic system. Pines and Slichter~\cite{ps55} have discussed magnetic relaxation by fluctuating fields in terms of a random walk approach and considered also the nuclear relaxation by conduction electrons in metal. Adopting their argumentation to the present case, the time $\tau_1$ is interpreted as 
$\tau_1=p(T)\tau_{\mathit{dwell}}$. This dwelling time $\tau_{\mathit{dwell}}=\hbar/E_F$ is roughly the time a conduction electron spends on a given atom, and $p(T)$ denotes the probability that during $\tau_{\mathit{dwell}}$ a nuclear spin flip occurs. For a degenerate Fermi gas, this probability is $p(T)=T/T_F$ where $T_F$ is the Fermi temperature. With these arguments we can express our parameter $a$ as
\begin{equation}
a=\frac{\hbar}{k_B {T_F}^2}
\end{equation}
which yields for the Fermi temperature $T_F=1050$ K. Moreover, we find then that the residence time on an atom is $\tau_{\mathit{dwell}}=7.3 \times 10^{-15}$ s. The rather low value~\footnote{For comparison, Harshman and Mills~\cite{harsh} found $T_F=2290\pm100$ K.} of $T_F$ indicates that the degeneracy is lifted. Instead of a well defined Fermi edge, the distribution is smeared out and the temperature dependence of the chemical potential becomes 
important, preventing a quantitative analysis without further information.

Adapting these arguments to the present case, it means that at low temperature all the spin-lattice relaxation rates (and that also for Cu) in \ybco{7} can be explained by scattering from quasiparticles within a kind of Fermi liquid model. The influence of the AFM spin correlations is however manifest in the temperature-dependent interaction  $\Vs{k}{\alpha}$ whose properties will be investigated in subsection \ref{sub:V}. \\

At high temperatures the data can no longer be explained by a Fermi-liquid behavior since the influence of the contribution $\tau_2$ grows. According to the fit values, we have $\tau_1=\tau_2$ at $420$ K. We defer a discussion of $\tau_2$ and the crossover $\tau_1$ to $\tau_2$ to later sections, until we have looked at more NMR/NQR data in other cuprates. \\
As concerns the ansatz (\ref{taumod}) for the effective correlation time we note that the interactions between the nuclei and the electronic system are the same whatever the ``basic'' relaxation mechanisms are. They are determined by the hyperfine energies $\Vs{k}{\alpha}(T)$ which, however, change from strong AFM correlations at $100$ K to weak AFM correlations at around $400$ K. Whether we really have two different ``basic'' relaxation mechanisms at work or whether they are two different manifestations of the same mechanism is an open question. \\
A possible explanation for the special form of (\ref{taumod}) is as follows. Adding two rates $\tau^{-1}_\mathit{short}$ and $\tau^{-1}_\mathit{long}$ means that the phase correlations between the processes associated with the longer time are destroyed by the impacts of those associated with the shorter time, since immediately after an impact of a process with time $\tau_\mathit{long}$ another impact of a process with the short time $\tau_\mathit{short}$ occurs. In our case, at low temperature $\tau_\mathit{short}\equiv \tau_1$ and $\tau_\mathit{long}\equiv \tau_2$, and vice-versa at high temperature. In addition, we also note that $\taueffm$ was originally introduced in Eq. (\ref{deftaueff}) in order to express the time dependence of $\langle S^{\alpha}_0 (t) S^{\alpha}_0 (0) \rangle$. If the time evolution is now governed by two processes so that the Hamiltonian is $\mathfrak{H}=\mathfrak{H}_1+\mathfrak{H}_2$, we get instead of Eq. (\ref{deftaueff}) that $ \langle S_0^{\alpha}(t) S_0^{\alpha}(0) \rangle \propto \mathrm{exp}\left( -|t|/\tau_1  -|t|/\tau_2 \right)$, provided $\mathfrak{H}_1$ and $\mathfrak{H}_2$ commute. We see that $\taueffm$ would thus be given by Eq. (\ref{taumod}).

\subsection{Comparison of model predictions and 
measurements at higher temperatures}\label{subsec:highT}
\begin{figure}
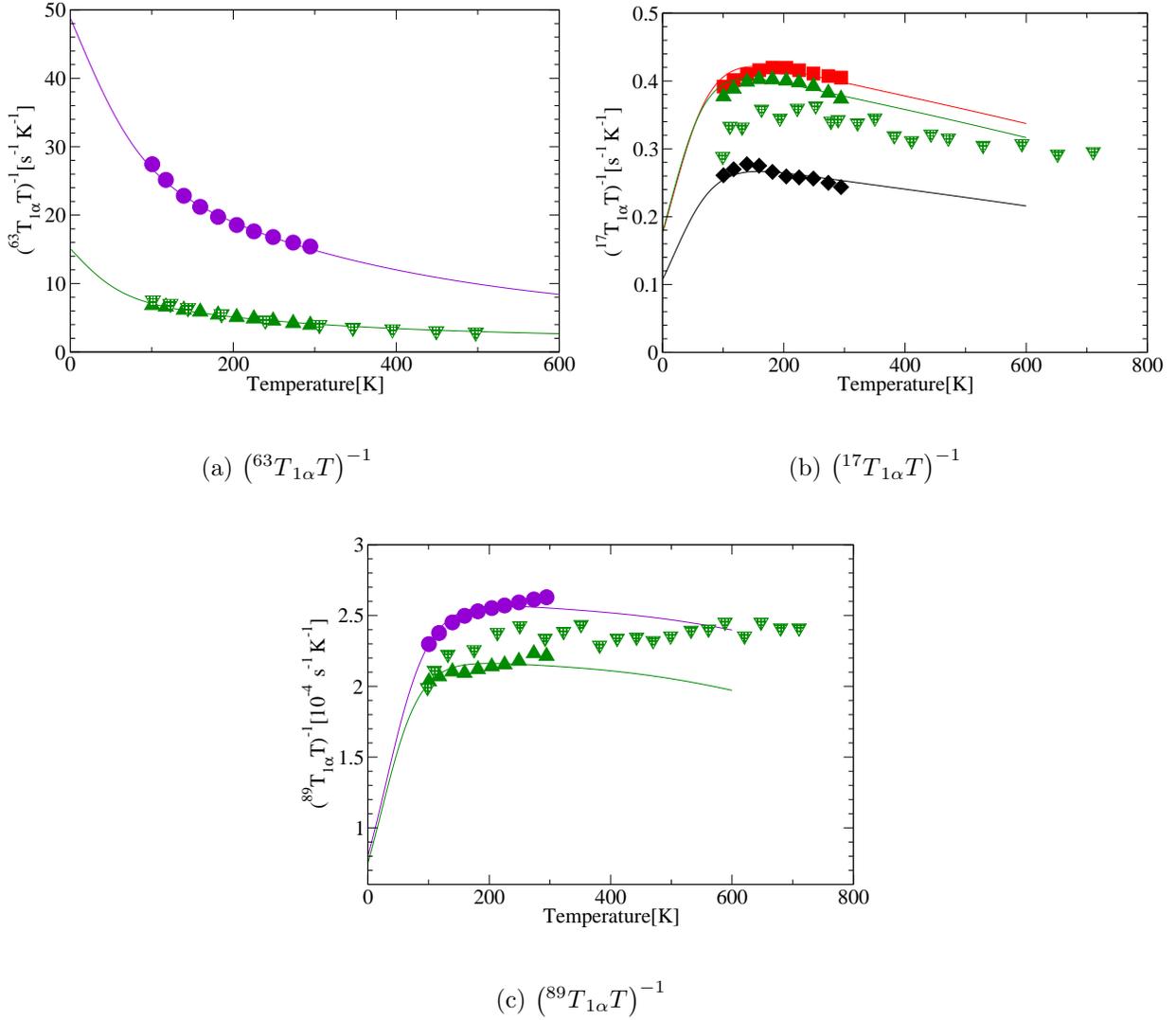

\begin{center}
\subfigure[$\TTm{63}{\alpha}$]{\label{fig:T1TCuex}
\includegraphics*[width=8cm]{T1TCuT_4ex.eps}}
\subfigure[$\TTm{17}{\alpha}$]{\label{fig:T1TOex}
\includegraphics*[width=8cm]{T1TOT_4ex.eps}}
\subfigure[$\TTm{89}{\alpha}$]{\label{fig:T1TYex}
\includegraphics*[width=8cm]{T1TYT_4ex.eps}}
\caption{Full symbols: original experimental data. Lines: fit as explained in the text. Hashed symbols: other data (see text). Direction 
$a$ (diamonds), $b$ (squares), $ab$ (circles) and $c$ (triangles). \label{fig:T1Tsex}}
\end{center}
\end{figure}

In \ybco{7} several spin-lattice relaxation rate measurements have been 
performed at temperatures well above the temperature interval which we 
used for fitting the model parameters.
So far we have not made use of these data since they do not allow us to 
extract the individual contributions $^kU_{\alpha}$. Now, however, it is of
interest to compare the extrapolated theoretical predictions with these data. 
We return to the customary $\TTm{k}{\alpha}$ representation and show in Fig. \ref{fig:T1TCuex} a plot of $\TTm{63}{\alpha}$ versus $T$.
The high-temperature data of 
Barrett {\it et al.}~\cite{barret:91} are denoted by triangles down, the original data~\cite{hammel:89,wals:88}  used for the
fit below room temperatures by full circles and full triangles and the model fit by the line.
In Fig. \ref{fig:T1TOex} we depict the temperature dependence of $\TTm{17}{\alpha}$ and include data points from 
Nandor {\it et al.}~\cite{nandor:99} who also reported values for $\TTm{89}{c}$ which are shown in Fig. \ref{fig:T1TYex}.

The predictions of the model obtained by extrapolation to higher 
temperatures are in excellent agreement with the measured relaxation rates of the copper. The downward trend of the relaxation rate of the oxygen is also well reproduced. Less good agreement is achieved in the case of the yttrium. The experimental rate seems to be more or less constant from $200$ K onwards, whereas we predict a slowly decreasing rate. This might be due to the simplistic model (Eq. {\ref{defUY}) we use, neglecting interplane spin correlations and dipolar couplings. We also note that there is some disagreement between the data of 
Nandor {\it et al.}\cite{nandor:99} and those of
Takigawa {\it et al.}\cite{taki:93} already in the temperature range below 
300 K.

On the whole however the model seems to explain the general trends of the data
well. A close inspection of the oxygen data, however, shows that the observed 
increase in the ratio of $\Tm{17}{a}/\Tm{17}{b}$ with decreasing temperature reported by Martindale~\cite{martind:98} is not reproduced.
The increase of the ratio $\Tm{17}{c}/\Tm{89}{c}$ for 
decreasing temperature observed by Takigawa {\it et al.}~\cite{taki:93} is 
qualitatively reproduced but lacks quantitative agreement. The ratio $\Tm{63}{c}/\Tm{17}{c}$ is however in very good agreement with our predictions and will be discussed in Section \ref{sec:ybco663}.

\subsection{Temperature dependence of interaction energies and Korringa relation}\label{sub:V}
It is instructive to discuss the calculated values $\Vs{k}{\alpha}(T)$ shown in Fig. \ref{fig:Utaus}.
\begin{figure}
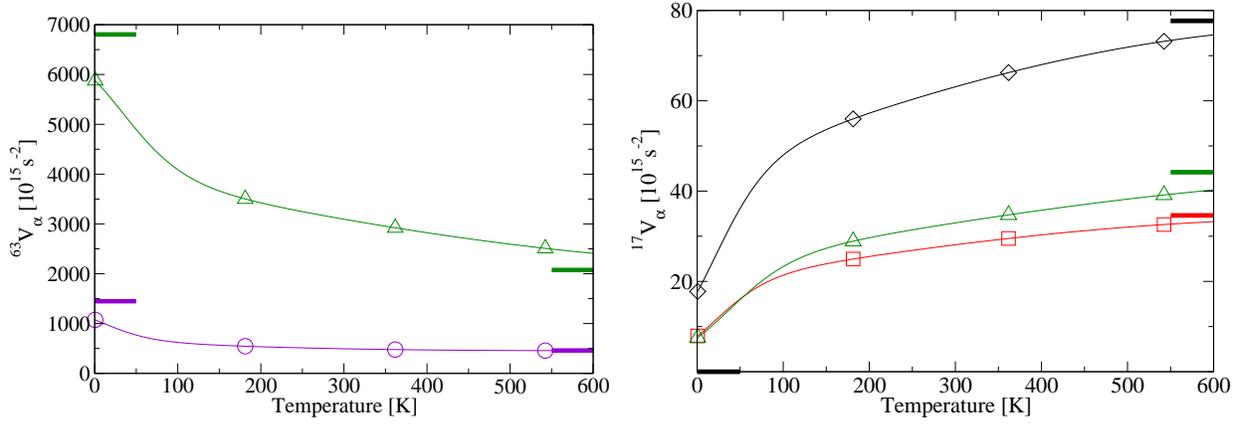
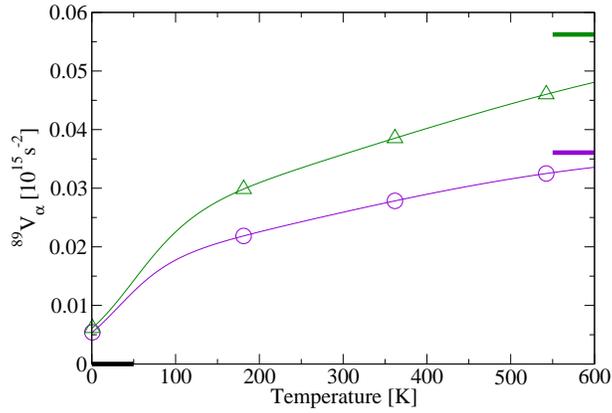

\begin{center}
\subfigure[$\Vs{63}{\alpha}$]{\label{fig:CutauUs}
\includegraphics*[width=8cm]{UtauCuT_4ex.eps}}
\subfigure[$\Vs{17}{\alpha}$]{\label{fig:OUtaus}
\includegraphics*[width=8cm]{UtauOT_4ex.eps}}
\subfigure[$\Vs{89}{\alpha}$]{\label{fig:YUtaus}
\includegraphics*[width=8cm]{UtauYT_4ex.eps}}
\caption{Temperature dependence of $\Vs{k}{\alpha}$ in the direction 
$a$ (diamonds), $b$ (squares), $ab$ (circles) and $c$ (triangles), which reflects the change in the degree of coherency. The bars at high 
temperatures indicate 
the values $\Vsn{k}{\alpha}$ with vanishing correlations, those at $T=0$ the fully correlated values $\Vfc{k}{\alpha}$.
\label{fig:Utaus}}
\end{center}
\end{figure}
The bars on the left in each figures indicate the extremal values $\Vsn{k}{\alpha}$ (Eqs. \ref{Vfc}) if the system was fully antiferromagnetic. The bars on the right show the other extreme $\Vsn{k}{\alpha}$  (Eqs. \ref{V0}), for a system with no correlations left. In the case of the 
oxygen (Fig. \ref{fig:OUtaus}), $\Vs{17}{\alpha}(T) = \Vsn{17}{\alpha} \left( 1-|K_{01}^{\alpha}(T)| \right) $ 
increases with $T$ for all directions $\alpha$ since the 
spin correlations tend to zero. For copper (Fig. \ref{fig:CutauUs}), $^{63}V_{\alpha}(T)$ drops 
with increasing temperature since the values
of the hyperfine interactions incidentally are such that the values for 
fully AFM correlations $\Vfc{63}{\alpha}$ (bars on the left)
are higher than for no correlations $\Vsn{63}{\alpha}$ (bars on the right), 
and that 
for both directions $\alpha$ parallel and perpendicular to the planes. The different
temperature dependencies of the rates measured at the copper and at the oxygen thus find a straightforward
explanation. While the ``basic'' relaxation rates are identical for both nuclei,$\Vs{63}{\alpha}(T)$ has an opposite temperature behavior as $\Vs{17}{\alpha}(T)$ due to the particular values of the hyperfine interaction energies, which are atomic properties.

The temperature dependencies of the rates $\us{k}{\alpha}$ result from a 
delicate balance between $\Vs{k}{\alpha}$ and $\taueffm(T)$. From 100 K
to 300 K, $\taueffm$ increases by about 170 \%, whereas $\Vs{63}{c}$ drops by
24 \% but $\Vs{17}{a}$ increases by 32 \%. Some $\Vs{k}{\alpha}$ values depend
crucially on the precise values of the hyperfine energies. Moreover, the 
rates $\Tm{k}{\alpha}$ which are directly accessible by experiments 
are linear combinations of the $\us{k}{\alpha}$. It is therefore important to study in details the interplay between $\taueffm(T)$ and $\Vs{k}{\alpha}(T)$.

We illustrate in Appendix \ref{app:V} the various contributions to $\Vs{63}{\alpha}(T)$ and discuss the interplay between nearest and
further distant AFM spin correlations.

At this point we just note that in the framework of the MMP model, the Fermi liquid contribution which dominates the relaxation rates of the oxygen and yttrium contains no correlations, i.e. would correspond to $\Vsn{k}{\alpha}$ (shown by the high temperature bars in Fig. \ref{fig:Utaus}).

It seems now also appropriate to comment on the Korringa relation
which has been discussed in numerous publications on NMR data
analysis in cuprates. The original Korringa relation has been derived for an
isotropic system. To get a similar relation for layered materials would
require: (i) temperature independent $\Vs{k}{\beta}$ and  $\Vs{k}{\gamma}$ and
 a  $\taueffm(T)$ which is proportional
to $T$ and to $\rho^2$, the square of the density of states at the
Fermi surface, (ii) a temperature independent static spin susceptibility in
direction $\alpha$, or (iii) an incidental cancellation of the corresponding
temperature dependencies in all these quantities which
in the frame of the present model and in view of Eq. 23 and Fig. 8 
is very unlikely.
From these considerations we conclude that looking for and discussing 
Korringa relations for NMR data in layered cuprates is probably a vain endeavor.

In Appendix~\ref{app:V} we show how an apparent linear temperature dependence 
of the oxygen relaxation rate may occur in a limited temperature interval.


\subsection{Spin-spin relaxation}\label{sub:spinspin}
It is worth at this stage to connect the present model for the spin-lattice relaxation in cuprates with the theory and experiments of the nuclear spin-spin relaxation rate $\Tg{}{}$ which have been put forward by Slichter and coworkers~\cite{penn:91,imaisliT2G:92,imaisliT2G:93,haase:99}. $\Tg{}{}$ measures the strengths of the indirect nuclear spin-spin interaction mediated by the non-local static spin susceptibility $\chir{}(\vec{r})$. 

Imai {\it et al.}~\cite{imaisliT2G:92} were the first to point out the importance of $\Tg{}{}$ to obtain information about the AFM exchange between the electron spins. They demonstrated that low field NMR measurements of $\Tg{63}{}$ give a strong quantitative constraint on $\chir{}(\vec{q})$ in cuprates. They discovered that the staggered susceptibility $\chir{}(\vec{Q})$ follows a Curie-Weiss law in \ybco{7}.

Since Haase {\it et al.}\cite{haase:99} also presented a derivation of $\chir{}(\vec{r})$ in direct space, it is of interest to compare Eq. (6) in Ref.~\onlinecite{haase:99} with our expression (\ref{VsdefCu}) for $\Vs{63}{c}(T)$. The functional form is the same but we note that $\Tg{63}{}$ requires the knowledge of the correlations at any distance $\vec{r}$, whereas $\Vs{63}{c}$ contains only the closest correlations. From the fit of the relaxation rate data we got values for $\Ks{01}{\alpha}$, $\Ks{12}{\alpha}$ and $\Ks{13}{\alpha}$ which, by assumption, could be provided by a static spin susceptibility
\begin{equation}\label{suscr}
\chir{\alpha}(\vec{r}) \propto \cos{(\vec{Q}\cdot \vec{r})}
e^{-r/\lambda_{\alpha}}.
\end{equation}
We have however no information on $\chir{\alpha}(\vec{r})$ for $|\vec{r}|<1$ and for $|\vec{r}|>2$. Nevertheless, the Fourier transform at $\vec{q}=\vec{Q}$ gives
\begin{equation}\label{chiQ}
\chir{\alpha}(\vec{Q})=2\pi\lambda_{\alpha}^{2} E \chir{\alpha}
(\vec{q}=0)
\end{equation}
where $E$ is an enhancement factor that, in the present approach, cannot be determined (expression \ref{chiQ} compares to the isotropic $\chir{}(\vec{Q})=\alpha \xi^2$ in the MMP model~\cite{mmp}). Assuming that Eq. (\ref{suscr}) also applies to correlations at arbitrary distances, we evaluate $\Tg{63}{}$ using the values $\lambda_c(T)$ obtained from the analysis of the $\Tm{k}{\alpha}$ (subsection \ref{ssec:fitting}).The resulting temperature dependence is shown in Fig. \ref{fig:T2G} (solid line) together with the data from Imai {\it et al.}~\cite{imaisliT2G:93} (circles).
\begin{figure}
\begin{center}
\includegraphics*[width=8cm]{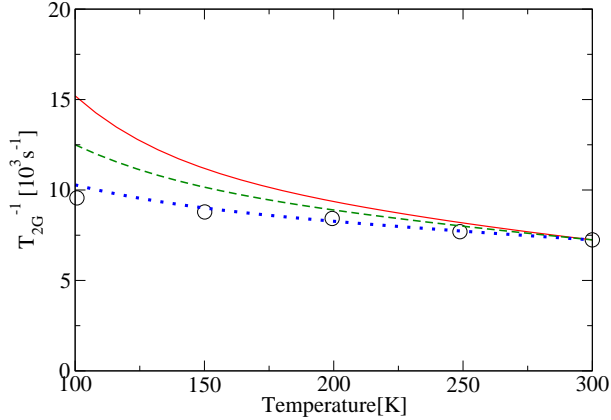}
\caption{\ybco{7}. Comparison with static results (circles) from Imai {\it et al.}~\cite{imaisliT2G:93}. All correlations included (full line), only $\Ks{01}{c}$,$\Ks{01}{c}$ and $\Ks{01}{c}$ (dashed line) and only NN correlations (dotted line). \label{fig:T2G}}
\end{center}
\end{figure}
$F:=E\chir{}(\vec{r}=0)/\mu_B^2$  was taken in (\ref{chiQ}) as a parameter to adjust, and we calculated it so that our result at $T=300$ K corresponds to the data. We found in this case $F=8.0\,\mathrm{eV}^{-1}$. Our prediction of the temperature dependence deviates from the measurements, which may indicate that our assumption of an exponential decay of $\chir{}(\vec{r})$ over-emphasizes the contributions of high order correlations. If we drop all the correlations except those coming into the $\Tm{k}{\alpha}$, that is $\Ks{01}{c}$, $\Ks{12}{c}$ and $\Ks{13}{c}$, we get the dashed line in Fig. \ref{fig:T2G} with $F=8.6\,\mathrm{eV}^{-1}$. A better correspondence with the data is obtained if we only keep   $\Ks{01}{c}$ (dotted line), where then $F=11.1\,\mathrm{eV}^{-1}$.\\

On the whole, these results are in agreement with those obtained by Imai {\it et al.}~\cite{imaisliT2G:92,imaisliT2G:93} who utilized a Gaussian form for the $q$-dependence of $\chir{}(\vec{q})$ near $\vec{q}=\vec{Q}$. In this work we have used a different parameterization of the AFM correlations (which yields shorter correlation lengths), but it seems worthwhile to test further the implications of the present model on $\chir{}(\vec{Q})$. An inspection of Eq. \ref{chiQ} shows that the increase of $\lambda_\alpha^2$ with decreasing temperature determines the increase of staggered susceptibility, provided that $E\chir{}(\vec{q}=0)$ is temperature independent. In Fig. \ref{fig:lambquad} we plotted the $\lambda^2_\alpha$ versus $1/T$ as we obtained them in subsection \ref{ssec:fitting} from the fit on all spin-lattice relaxation data.
\begin{figure}
\begin{center}
\includegraphics*[width=8cm]{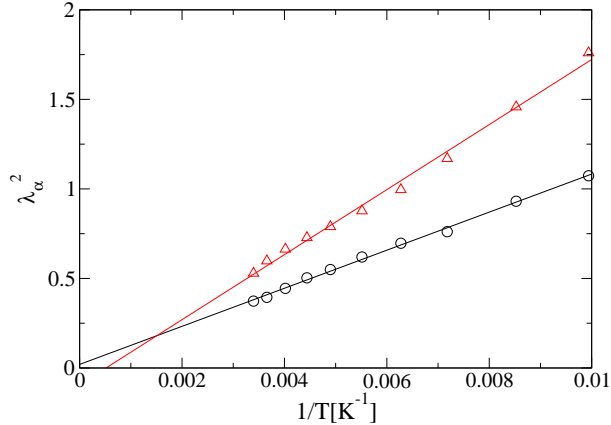}
\caption{$\lambda_\alpha^2$ versus $1/T$ in the direction ab (circles) and c(triangles). Lines are averages. \label{fig:lambquad}}
\end{center}
\end{figure}
It is astonishing how well these values obey a Curie (or Curie-Weiss) behavior. It seems that upon cooling the spins in \ybco{7} remember their tendency to align antiferromagnetically, but they are prevented from doing so as the more energetically-favored superconductivity sets in. This feature has already been observed and discussed by Imai {\it et al.}~\cite{imaisliT2G:92,imaisliT2G:93}. \\
An important question at this stage is whether those AFM correlations persist in the superconducting state. There are strong indications that the correlations develop an extreme anisotropy~\cite{uldryAni} below $T_c$. The question of the persistence of correlations in this state will reveal extremely interesting information about the interplay of magnetism and superconductivity which, however, is not the focus of the present work.



\section{Underdoped YBCO compounds}\label{sec:underdop}
In this section the model is applied to analyze NMR and NQR experiments in 
YBa$_2$Cu$_3$O$_{6.63}$ and in YBa$_2$Cu$_4$O$_8$. We do not have for these compounds the same full set of data as for the optimally doped YBa$_2$Cu$_3$O$_7$, and we 
therefore restrict our analysis to exploring the general trends of the doping and 
temperature dependencies of the AFM correlations and the effective 
correlation time $\taueffm$. We retain the same hyperfine interaction energies which have been
determined for YBa$_2$Cu$_3$O$_7$ in the analysis of YBa$_2$Cu$_3$O$_{6.63}$
and even for YBa$_2$Cu$_4$O$_8$. In the latter case, we are able to compare our model predictions with high temperature measurements.

\subsection{\ybco{6.63}}\label{sec:ybco663}
To determine the rates $\us{k}{\alpha}$ we used the following published
spin-lattice relaxation data: the copper data from Takigawa {\it et al.}~\cite{taki:91} (only providing $\Tm{63}{c}$), the oxygen data
$\Tm{17}{\alpha}$ taken from Martindale {\it et al.}~\cite{martind:98}, and the yttrium
data $\Tm{89}{\alpha}$ from Takigawa {\it et al.}~\cite{taki:93}. They cover the temperature range from $T_c$ up
to room temperature. We are not aware of measurements at elevated temperatures.
The resulting relaxation rates $\us{k}{\alpha}$ are shown (symbols) in Fig. \ref{fig:Usex6}. The lines in this figure will be explained later.
\begin{figure}
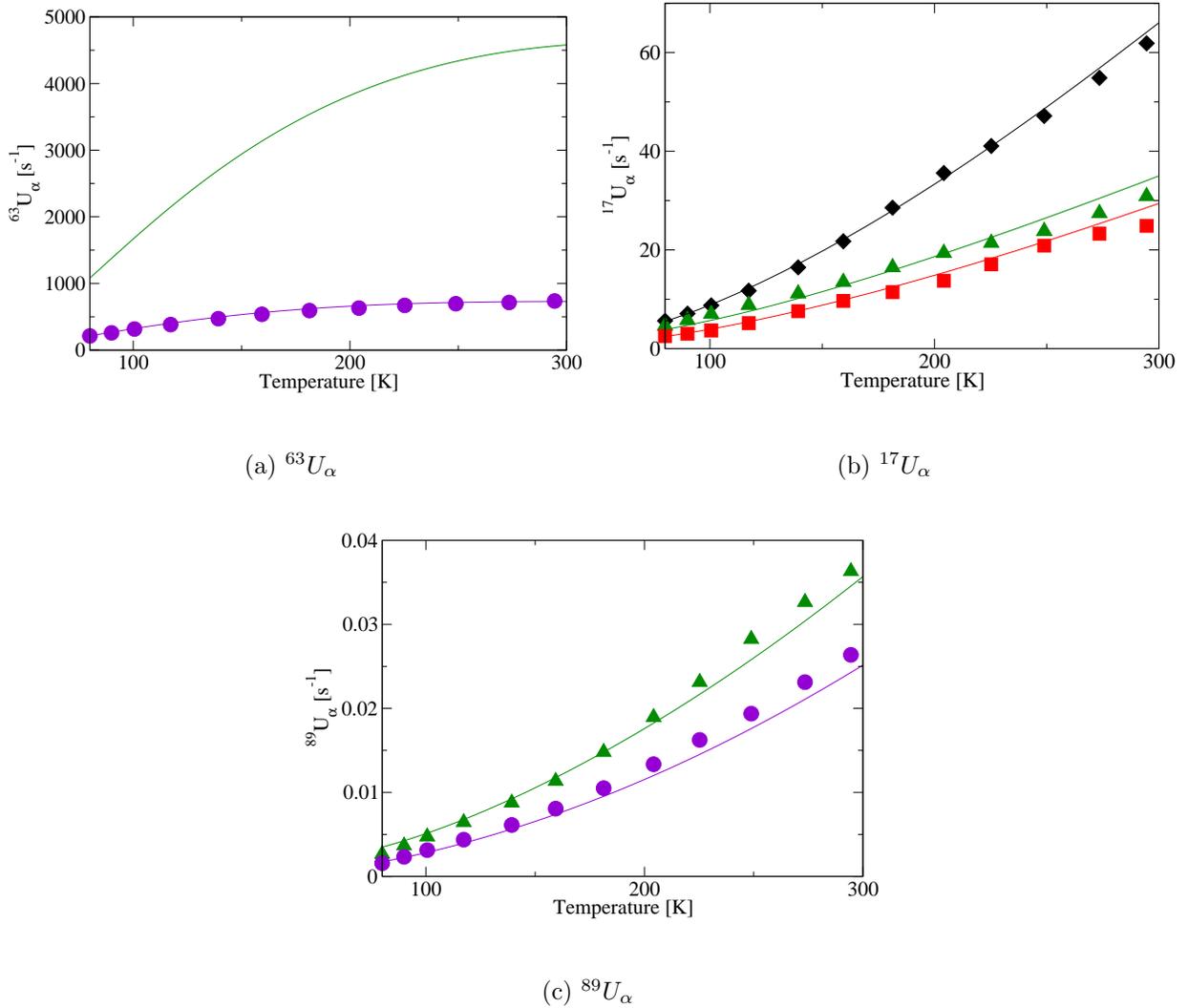

\begin{center}
\subfigure[$\us{63}{\alpha}$]{\label{fig:UsCuex6}
\includegraphics*[width=8cm]{Us63Uex4_Y6.eps}}
\subfigure[$\us{17}{\alpha}$]{\label{fig:UsOex6}
\includegraphics*[width=8cm]{Us17Uex4_Y6.eps}}
\subfigure[$\us{89}{\alpha}$]{\label{fig:UsYex6}
\includegraphics*[width=8cm]{Us89Uex4_Y6.eps}}
\caption{\ybco{6.63}. Symbols are experimental data, lines the fit as explained in the text. Directions: $a$ (diamonds), $b$ (squares), $ab$ (circles)and 
$c$ (triangles). \label{fig:Usex6}}
\end{center}
\end{figure}
A comparison 
with those of the optimally doped material (Fig. \ref{fig:Us}) reveals that now the temperature 
dependencies deviate more strongly from a linear behavior with a convex (concave)
curvature for $\us{63}{\alpha}(T)$ $\left(\us{17}{\alpha}(T) \right)$, but we do not consider this to be a dramatic contrast.

The same fitting procedure explained in \ref{ssec:fitting} was then carried out except that the 
values for the hyperfine interaction energies were kept fixed at the optimized values found for \ybco{7} (Table \ref{tab:hfeV}). It is therefore expected that the quality of the fit will be less good than for \ybco{7}. The result is depicted in Fig. \ref{fig:TUlamb4ex}, which shows the temperature dependence of the correlation lengths  $\lambda_{\alpha}$ compared with the values for the optimally doped material.
\begin{figure}
\begin{center}
\subfigure[$\lambda_{ab}$ (circles) and $\lambda_{c}$ (triangles). \ybco{7} (full lines), \ybco{6.63}(dotted lines). ]{\label{fig:TUlamb4ex}
\includegraphics*[width=8cm]{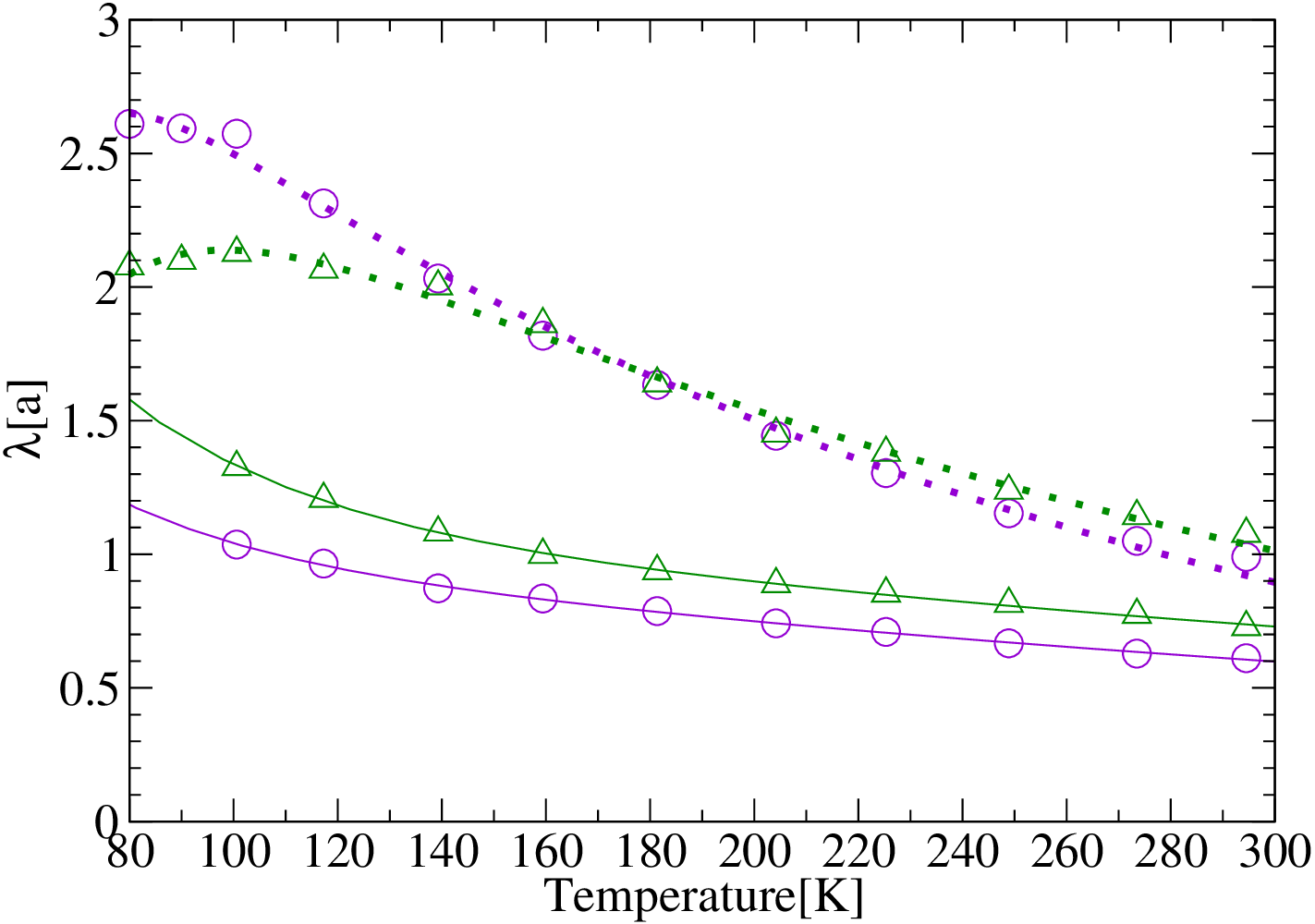}}
\subfigure[$K_{01}^{ab}$ (circles), $K_{01}^{c}$ (triangles)
in \ybco{6.63}.]{\label{fig:Ucorrel4}
\includegraphics*[width=8cm]{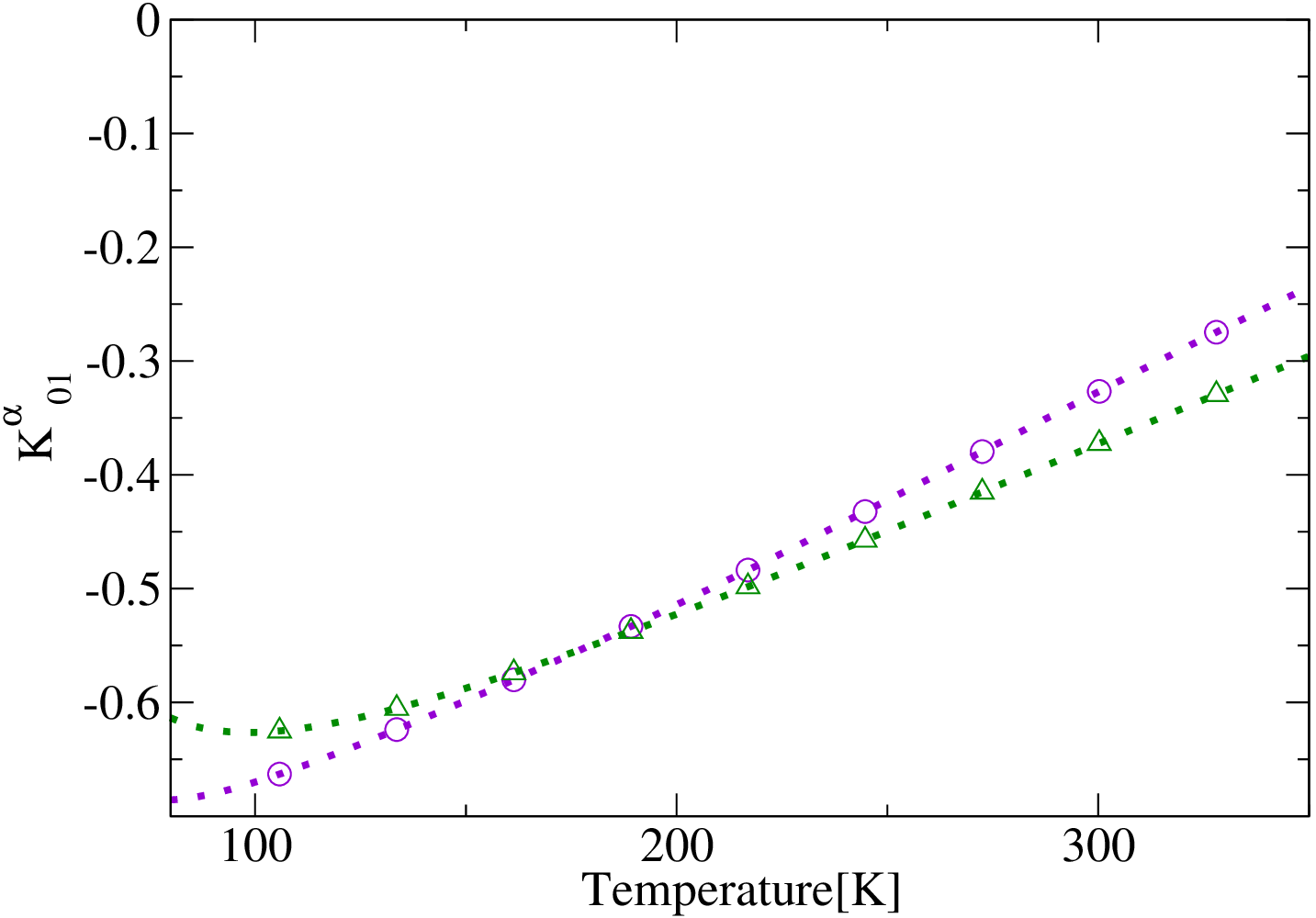}}
\caption{Correlation lengths and correlations in \ybco{6.63}. \label{fig:lambcorrU4ex}}
\end{center}
\end{figure}
As expected, the values for  $\lambda_{\alpha}$ are higher 
(by about a factor 2) in the underdoped compound but exhibit a peculiar 
crossover when the temperature drops below 100 K. In this region we also get
 $\lambda_{c} < \lambda_{ab}$. Whether this behavior is physically really significant
or just shows the inadequacy of the postulated model is at the moment open. As was done for \ybco{7}, the correlation lengths can be fitted with a sum of exponential functions. The correlations built using the exponential 
fit are plotted in 
Fig. \ref{fig:Ucorrel4}. 
For completeness we have plotted in Fig. \ref{fig:Utaus6} the 
calculated $\Vs{k}{\alpha}$.
\begin{figure}
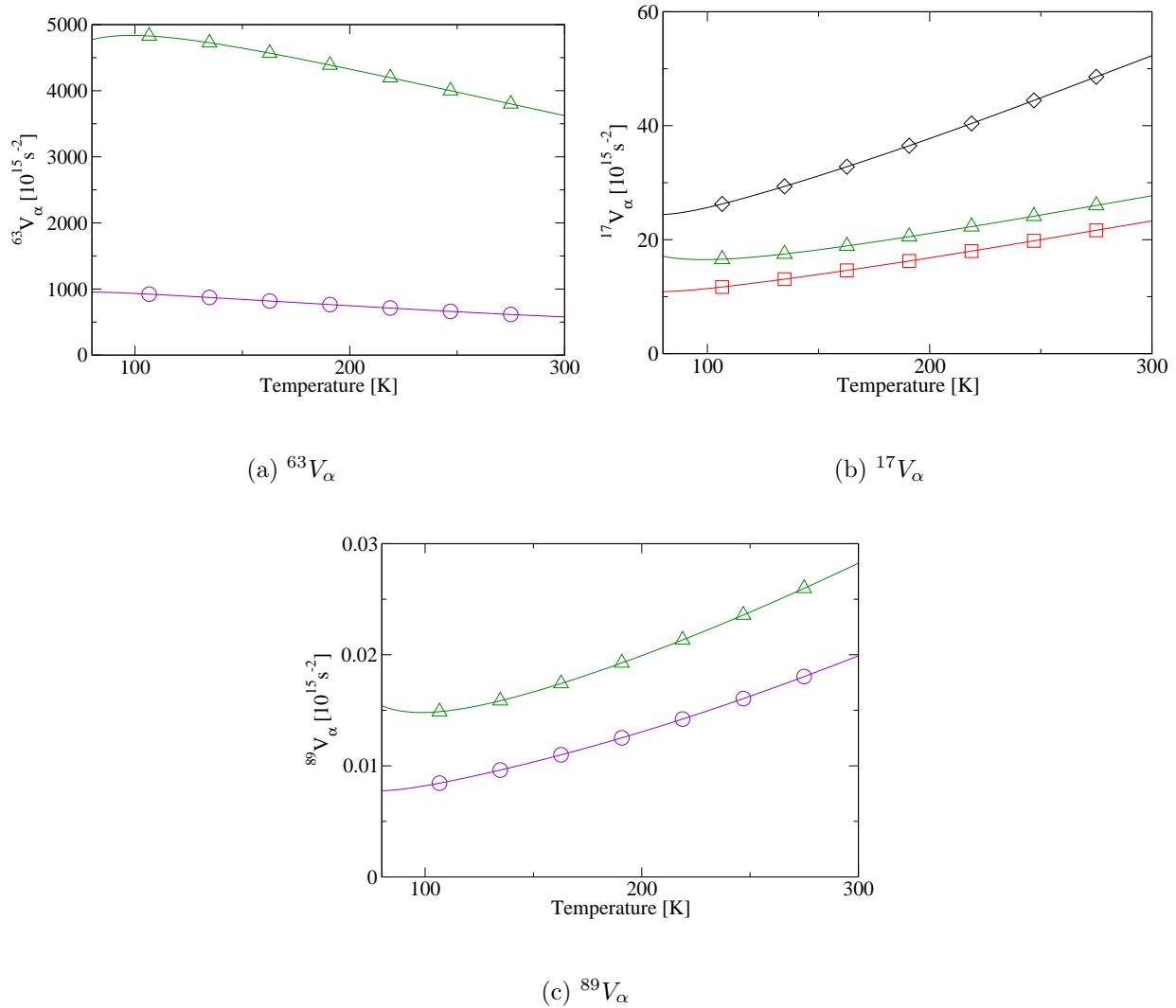

\begin{center}
\subfigure[$\Vs{63}{\alpha}$]{\label{fig:CutauUs6}
\includegraphics*[width=8cm]{UtauCuU_4ex.eps}}
\subfigure[$\Vs{17}{\alpha}$]{\label{fig:OUtaus6}
\includegraphics*[width=8cm]{UtauOU_4ex.eps}}
\subfigure[$\Vs{89}{\alpha}$]{\label{fig:YUtaus6}
\includegraphics*[width=8cm]{UtauYU_4ex.eps}}
\caption{\ybco{6.63}. Calculated $\Vs{k}{\alpha}$, in the directions $a$ (diamonds), $b$ (squares), $ab$ (circles) and $c$ (triangles). \label{fig:Utaus6}}
\end{center}
\end{figure}

The consistency check in analogy with \ref{sub:consistency} gives 6 different 
values for $\taus{k}{\alpha}(T)$ which are gathered in Fig. \ref{fig:tauU4det}. Although this time the spread 
among the different values is considerably larger than that for \ybco{7}
(see Fig. \ref{fig:tauT4det}) the
agreement is still surprisingly good.
\begin{figure}
\begin{center}
\includegraphics*[width=8cm]{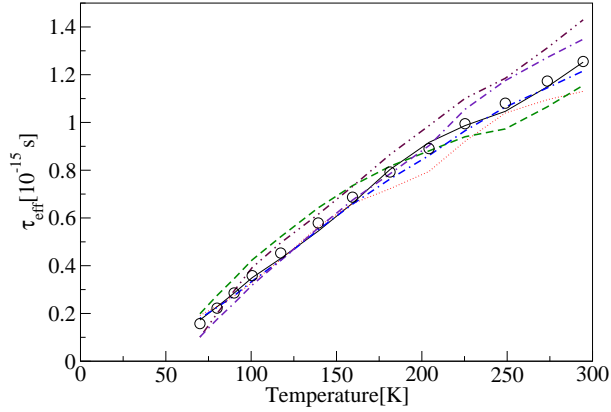}
\caption{ $\taueffm$ in \ybco{6.63} obtained from a $\lambda$ minimization at the optimized parameters (Table \ref{tab:hfeV}). The circles are the calculated average. 
$\taus{17}{a}$ (thin full line), 
 $\taus{17}{b}$ (thin dotted), 
 $\taus{17}{c}$ (thin dashed), 
 $\taus{63}{ab}$ (thick dash-dotted), 
 $\taus{89}{ab}$ (thick dash-dot-dot-dashed), 
 $\taus{89}{c}$ (thick dot-dash-dash-dotted). \label{fig:tauU4det}}
\end{center}
\end{figure}
The temperature dependence of the
calculated mean values (circles in Fig. \ref{fig:tauU4det}) could not be fitted with the 
function (\ref{tau1nogap}). However, the same ansatz
\begin{subequations}
\begin{gather}\label{tausgap}
\taueffm^{-1} = \tau_{1}^{-1} +  \tau_{2}^{-1} \\
\textrm{but with} \quad
\tau_1 = a T e^{-g/T} \quad \textrm{and}\quad  \tau_{2} = \textrm{constant}\label{taumodgap}
\end{gather}
\end{subequations}
provides a very good fit (as is shown in Fig. \ref{fig:tauU4}) with the values 
a = 10 $\times 10^{-18}$ s/K, g= 97 K, and 
$\tau_2 = 2.9 \times 10^{-15}$ s.
\begin{figure}
\begin{center}
\includegraphics*[width=8cm]{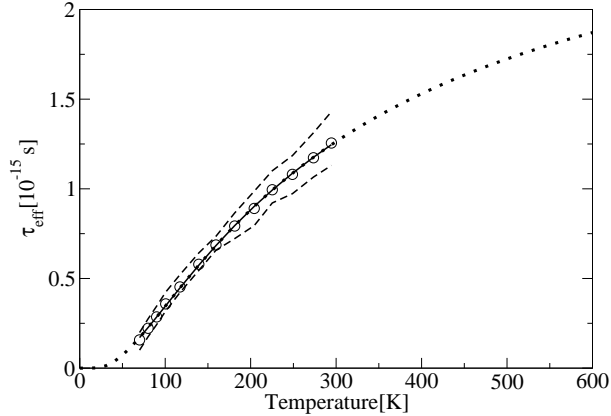}
\caption{ $\taueffm$ in $\mathrm{Y} \mathrm{Ba}_2 \mathrm{Cu}_3 
\mathrm{O}_{6.63}$ obtained from a $\lambda$ minimization at the optimized parameters. The circles are the calculated average, and the dashed lines the upper and lower bound. The line is the fit as explained in the 
text. \label{fig:tauU4}}
\end{center}
\end{figure}

The lines in Fig. \ref{fig:Usex6} show the model-calculated $U$s compared with the experiments, whereas Fig. \ref{fig:T1Tsex6} depicts the data and the fit in the usual $\TTm{k}{\alpha}$ versus $T$ representation.
\begin{figure}
\begin{center}
\subfigure[$\TTm{63}{\alpha}$]{\label{fig:T1TCuex6}
\includegraphics*[width=8cm]{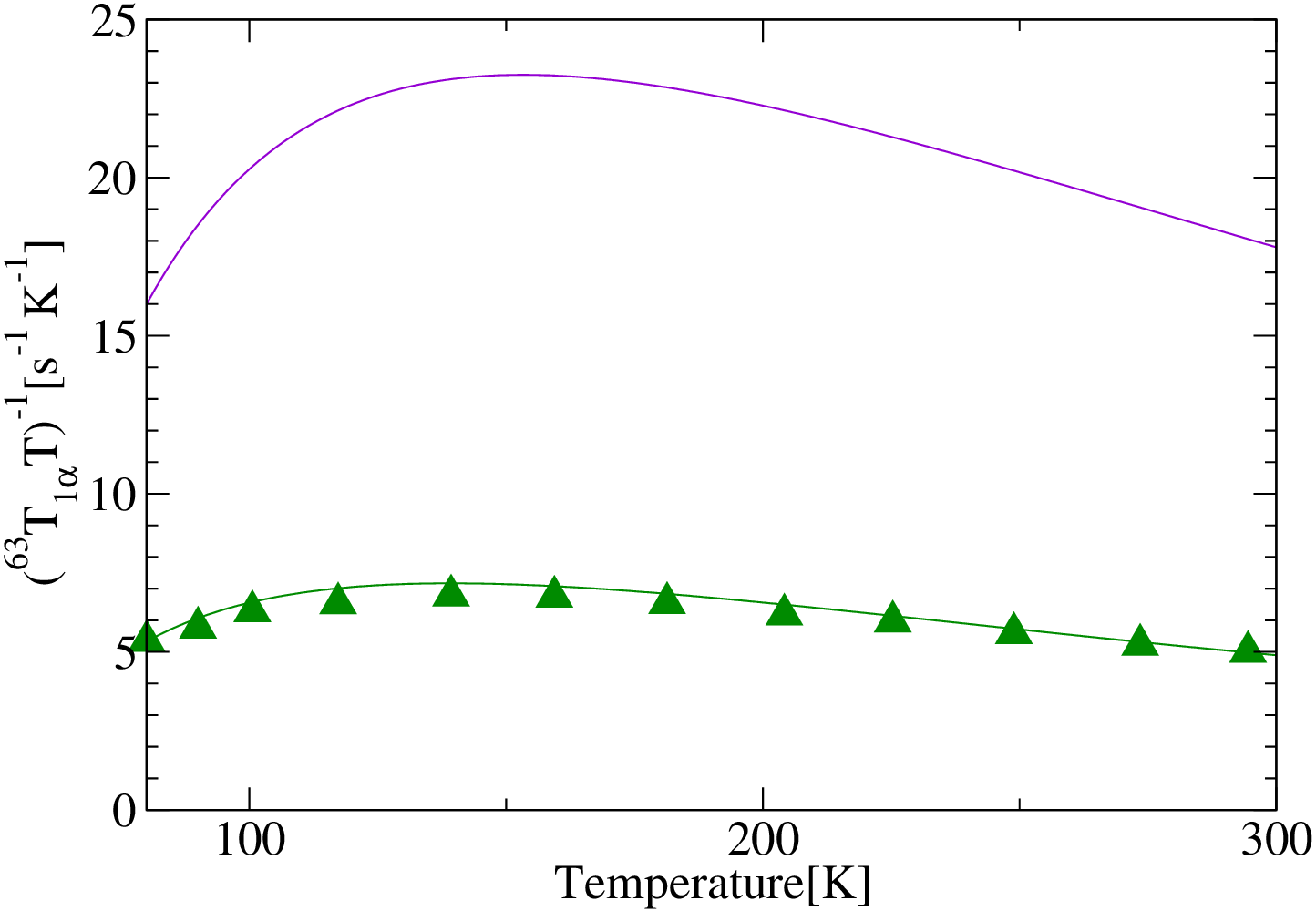}}
\subfigure[$\TTm{17}{\alpha}$]{\label{fig:T1TOex6}
\includegraphics*[width=8cm]{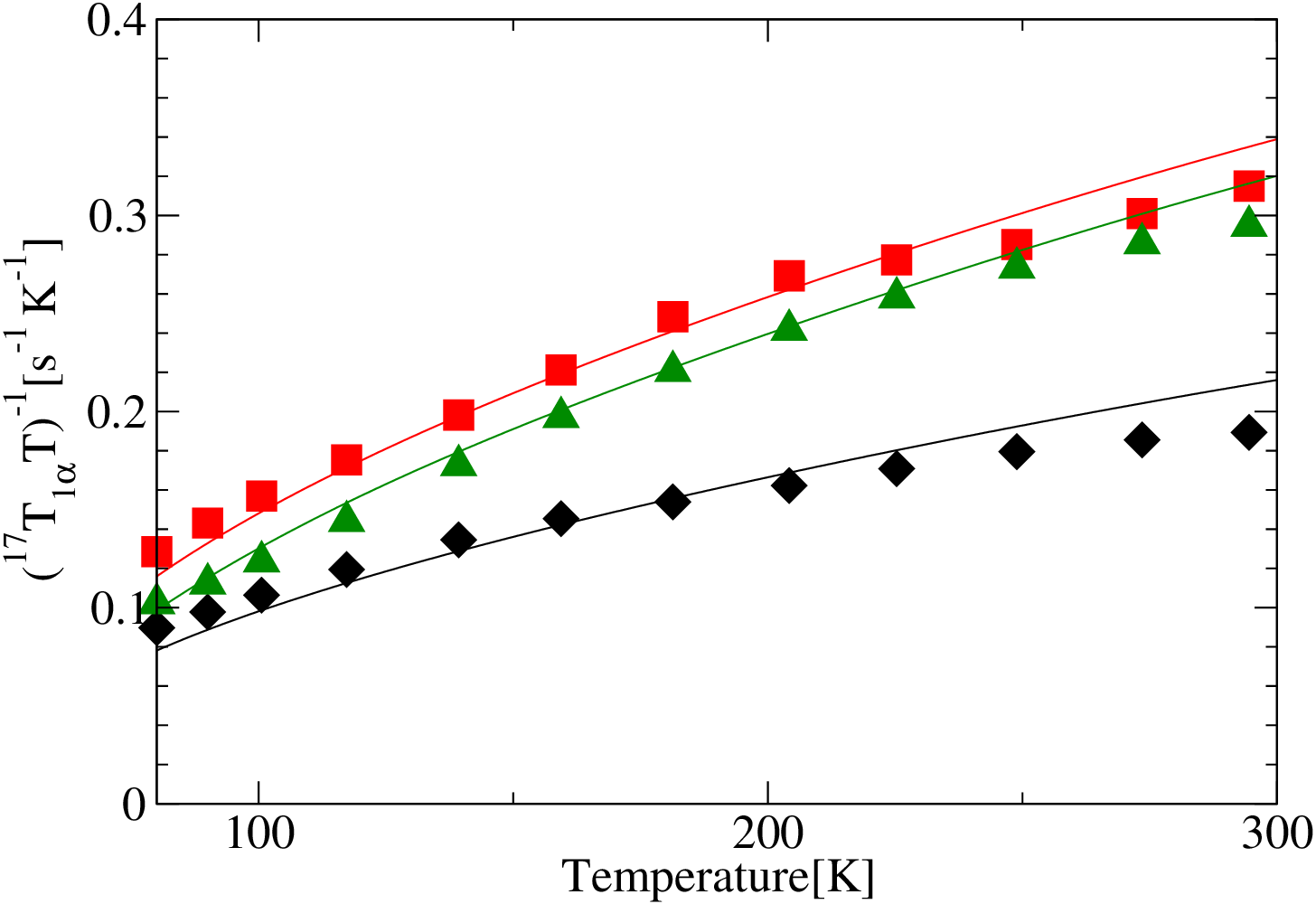}}
\subfigure[$\TTm{89}{\alpha}$]{\label{fig:T1TYex6}
\includegraphics*[width=8cm]{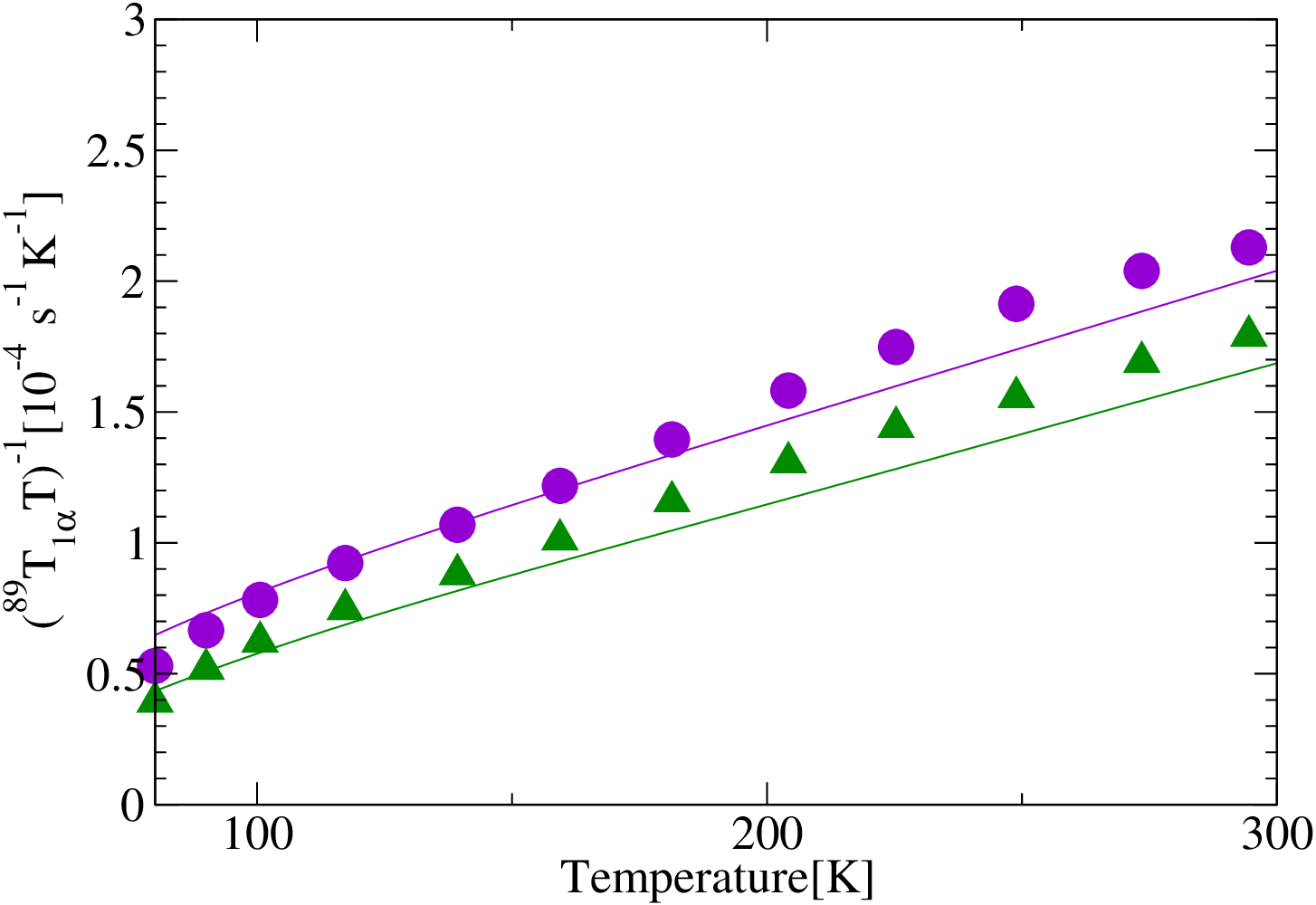}}
\caption{\ybco{6.63}. Symbols are experimental data, lines the fit as explained in the text. Directions: $a$ (diamonds), $b$ (squares), $ab$ (circles)and 
$c$ (triangles).\label{fig:T1Tsex6}}
\end{center}
\end{figure}
The agreement is now only approximate but could be improved by further adjustments (in particular the hyperfine field constants) and
weighting of the data according to their precision. A further confirmation that the fit is however convincing is given in Fig. \ref{fig:R63R17}, where we have plotted the ratio $\Tm{63}{c}/\Tm{17}{c}$. 
\begin{figure}
\begin{center}
\includegraphics*[width=8cm]{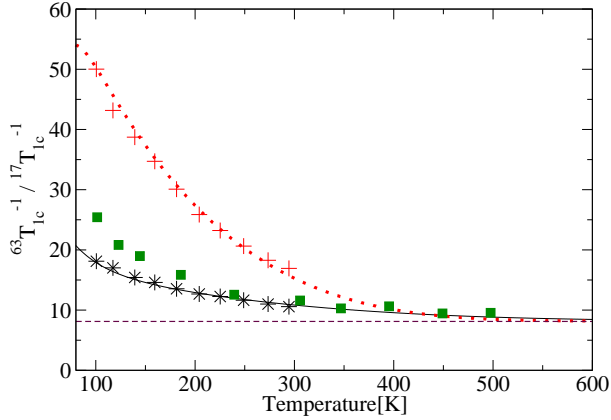}
\caption{$\Tm{63}{c}/\Tm{17}{c}$ in \ybco{7} (stars *) and \ybco{6.63} (plusses +). Symbols are experimental data as cited in the text. The full and dotted lines are the predictions derived from the model, and the constant dashed line is the expected value in the absence of AFM correlations.  \label{fig:R63R17}}
\end{center}
\end{figure}
On top of our original data (stars and plusses) and our model predictions (full and dotted line), we also show the data (squares) obtained by combining the high temperature \ybco{7} measurements of Barrett's {\it et al.}~\cite{barret:91} and Nandor's {\it et al.}~\cite{nandor:99} already discussed in \ref{subsec:highT}. At high temperature, we expect the AFM correlations to become negligible. In this case, the hyperfine fields must be fully uncorrelated and the ratio $\Tm{63}{c}/\Tm{17}{c}$ tends to $(A_{ab}^2+4B^2)/(C_a^2+C_b^2)$, which is about $8.2$. This value is marked by a dotted line in Fig. \ref{fig:R63R17}. As can be seen in this figure, the high temperature data confirms this prediction, which is a known result for the relaxation by randomly fluctuating hyperfine fields.

\subsection{YBa$_2$Cu$_4$O$_8$}
In stoichiometric YBa$_2$Cu$_4$O$_8$, NMR and NQR lines are much narrower than
in other cuprates and allow precise measurements. Raffa {\it et al.}~\cite{raffa} reported high accuracy $^{63}$Cu NQR 
spin-lattice relaxation measurements on $^{16}$O and  $^{18}$O
exchanged samples of YBa$_2$Cu$_4$O$_8$. They analyzed their data with the help
of the phenomenological relation
\begin{equation}\label{raf}
\TTm{63}{c}= C T^{-a} 
\left[ 1 - \tanh^2\left(\frac{\Delta}{2T}\right) \right]
\end{equation}
which worked reasonably well. To investigate a possible isotope
effect of the spin gap parameter $\Delta$ they could considerably improve the 
agreement between data and fit function by slightly adjusting the
temperature dependence of the function (\ref{raf}).

In terms of our model the temperature dependence of the relaxation rate is given by
\begin{equation}\label{raf2}
\Tm{63}{c} =  2 \:\us{63}{ab}(T) = 2 \:  \Vs{63}{ab}(T) \: \taueffm(T) .
\end{equation}
To test the quality of the ansatz (\ref{tausgap})-(\ref{taumodgap}) for the combined effective correlation time
we simply take the values 
for the $\Vs{63}{ab}(T)$ which we have extracted for the \ybco{7} system. 
Eq. (\ref{raf2}) is then used to determine the $\taueffm(T)$ from the data of Raffa {\it et al.}~\cite{raffa}. The data (for the $^{16}$O sample) and the fit are 
shown in Fig. \ref{fig:raffa_nurfit}. Fig. \ref{fig:raffa} shows the same data with the calculated values extrapolated outside the range 
(100 K $< T <$ 310 K), with the addition of high-temperature data from Curro {\it et al.}~\cite{curro} and Tomeno {\it et al.}~\cite{tomeno:94}. The temperature dependence of $\taueffm$ is represented in Fig. \ref{fig:rafftau}.
\begin{figure}
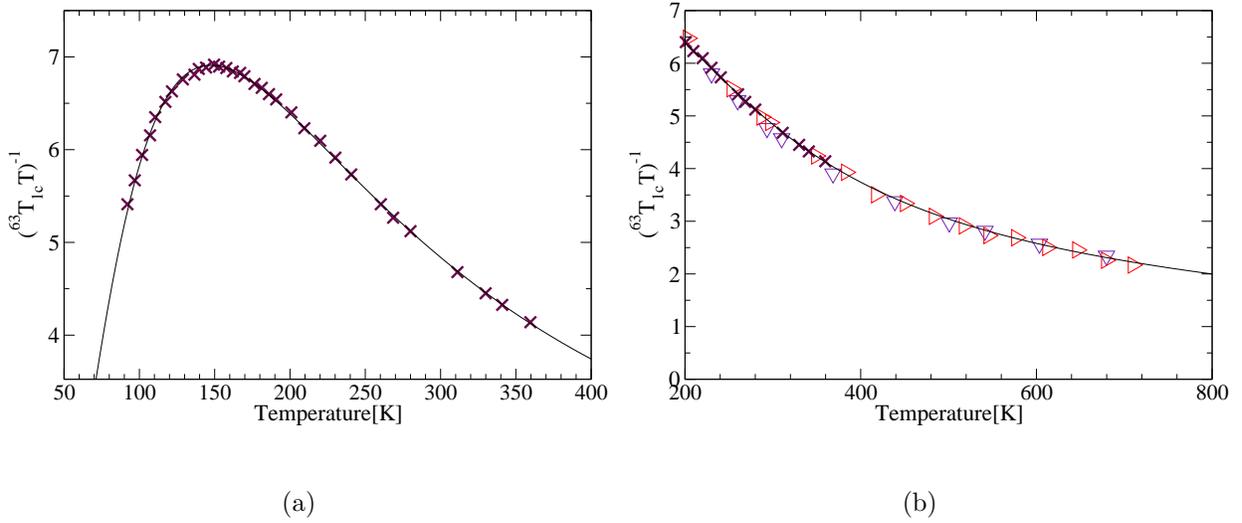

\begin{center}
\subfigure[]{\label{fig:raffa_nurfit}
\includegraphics*[width=8cm]{raffaTTCuc8_nurfit.eps}}
\subfigure[]{\label{fig:raffa}
\includegraphics*[width=8cm]{raffaTTCuc8.eps}}
\caption{(a): data from Raffa {\it et al }~\cite{raffa}(crosses) and fit as in the text (solid line). (b): Extrapolation of fit to higher temperatures and data from Curro {\it et al.}~\cite{curro} (triangles right) 
and Tomeno {\it et al.}~\cite{tomeno:94} (triangles down). \label{fig:raffas}}
\end{center}
\end{figure}
The fit gives $a =44 \times 10^{-18}$ s/K,  
g = $195$ K and $\tau_2$ = $1.9 \times 10^{-15}$ s.
Note that these values will change if the $^{63}V_{ab}(T)$ values appropriate to
 YBa$_2$Cu$_4$O$_8$ (which are expected to be somewhat larger between 100 K and 300 K due to enhanced correlations) can be  extracted.
\begin{figure}
\begin{center}
\includegraphics*[width=8cm]{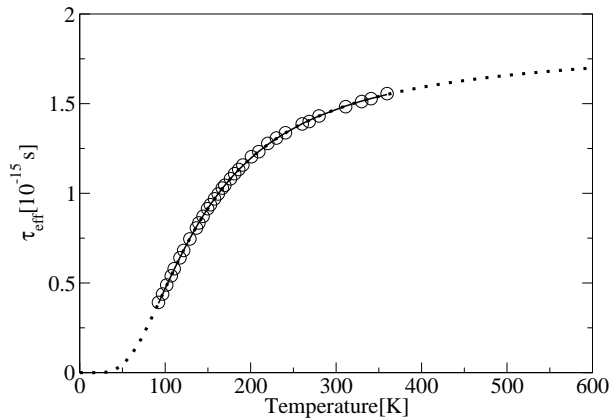}
\caption{$\taueffm$ (circles) is obtained from $\Vs{63}{ab}(T)$ and the data from Raffa {\it et al.}~\cite{raffa}. Dotted line: fit as in the text. \label{fig:rafftau}}
\end{center}
\end{figure}
We guess that our ansatz (\ref{tausgap}) and (\ref{taumodgap}) might 
find a better physical
motivation than (\ref{raf}). Moreover the fit is excellent and the extrapolation to higher temperatures is in addition in good agreement with the 
data available.


\section{Lasco compounds}\label{sec:lasco}


A large amount of NMR and NQR data exist also for La$_{2-x}$Sr$_x$CuO$_4$ for various
doping levels $x$. Mostly, relaxation rates have been reported 
for $\Tm{63}{c}$ with only few measurements on the oxygens. Haase {\it et al.}~\cite{haase:02} have been able to determine $K_{01}(T)$ ($\rho_{01}(T)$ in their work) from the linewidth of apical, planar O and Cu in La$_{2-x}$Sr$_{x}$CuO$_{4}$. For the optimally doped compound they find that the correlation decreases with the temperature and is about $-0.4$ at room temperature. The correlations are of course expected to be large for low doping. In the framework of our model we are not in a position to determine the correlation lengths $\lambda_{ab}$ and $\lambda_c$ since we do not have enough data. We therefore restrict ourselves in this section to a more qualitative discussion of the temperature dependence of $\Tm{63}{c}$. In particular, we can compare the high temperature limit of the effective correlation time $\tau_2$ to the relaxation mechanism of local magnetic moments in a paramagnet.

We reproduce in Fig. \ref{fig:lascoohsugi} the measurements of
Ohsugi {\it et al.}~\cite{ohsugi}
who present their data in a $\Tone{63}{c} T$ versus $T$ plot. 
\begin{figure}
\begin{center}
\includegraphics*[width=8cm]{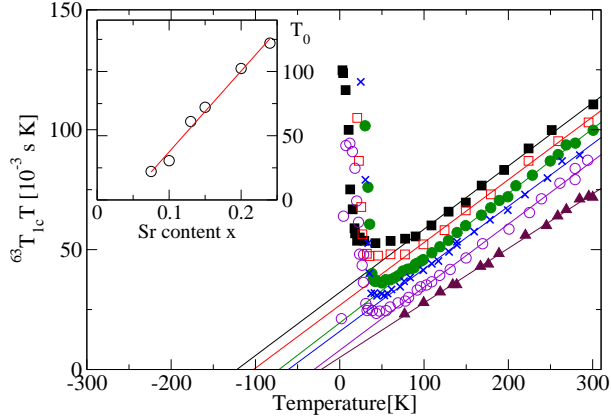}
\caption{$\Tone{63}{c}T$ vs $T$ plot with the data from 
Ohsugi {\it et al.}~\cite{ohsugi} for
\lasco. x=0.24 (full squares), x=0.2 (empty squares), x=0.15 (full circles), x=0.13 (crosses), x=0.1 (empty circles), x=0.075 (full triangles). In the inset, $T_0$ is plotted versus the strontium concentration. \label{fig:lascoohsugi}}
\end{center}
\end{figure}

The straight lines that
fit the data well at temperatures above 100 K correspond to
\begin{equation}
\Tone{63}{c} T = \alpha + \alpha T / T_0
\end{equation}
and led, in analogy to the Curie-Weiss law for the susceptibility in insulating
antiferromagnets, to the notion of ``Curie-Weiss behavior 
of  $\TTm{63}{c}$''.

In our model we also expect to find for La$_{2-x}$Sr$_x$CuO$_4$ a $\taueffm$ as given by (\ref{tausgap}) and (\ref{taumodgap}), that is
\begin{equation}
\Tone{63}{c} T = \frac{1}{2\, \Vs{63}{ab}(T)} \left[ \frac{1}{a} e^{g/T} + \frac{T}{\tau_2}\right]. 
\end{equation}
At high temperatures the AFM correlations vanish, 
$^{63}V_{ab}$ is 
temperature independent and $T \gg g$. We find indeed in this case a 
linear $T$ dependence and can identify
\begin{equation}
\alpha \equiv \frac{1}{2\, \, \Vsn{63}{ab}\, a} \quad \textrm {and}\quad  T_0 \equiv \frac{\tau_2}{a}.
\end{equation}
It is astonishing, however, that this linear temperature dependence dominates 
already at 
T~$\gtrsim 100$ K. It is probable that like in \ybco{7}, the temperature dependence of $\Vs{63}{ab}$ is weak. We have shown in Appendix \ref{app:V} that due to the incidental values of the hyperfine energies for copper, a cancellation of contributions from NN and further distant correlations occurs. 

Imai et al.~\cite{imaisli:93} have studied La$_{2-x}$Sr$_x$CuO$_4$ by NQR up to high temperatures
and found two striking results. Firstly, at high 
temperatures ($\sim$ 600 K) the 
$\Tm{63}{c}$ rates of all samples (with doping levels $x$= 0, 0.02, 0.04, 0.045, 0.075, and
0.15) were nearly identical although the system with $x$ = 0 is an insulator and its magnetic
behavior can well be described by a two-dimensional Heisenberg antiferromagnet, whereas the
samples with $x$ = 0.075 and 0.15 are conducting and, at low $T$ even superconducting.
Secondly, they observed that at high temperatures $\Tone{63}{c}$ becomes independent 
of temperature. 

The first striking result of Imai {\it et al.}~\cite{imaisli:93}, that is the nearly identical values 
of $\Tm{63}{c}$ for $x$=0 up to $x$=0.15, is a strong argument for identifying
$\tau_2$ as a time scale dictated by the dynamics of the antiferromagnetism which persists
well into the doped and overdoped region. This can be illustrated as follows. Moriya~\cite{moriya:56} has calculated $\Tm{}{}$ for a nucleus of a magnetic ion in an {\it insulating} antiferromagnet. In the paramagnetic regime he treated the exchange interaction within the model of a Gaussian random process with correlation frequency $\omega_e$. This was adapted by Imai {\it et al.}~\cite{imaisli:93} to a square lattice and appropriate hyperfine interaction energies for pure \lacuo with the result (equation (4) in Ref. \onlinecite{imaisli:93})
\begin{equation}
\Tm{63}{\infty}=\sqrt{2\pi}\left(A_{ab}^2+4B^2 \right)\frac{1}{4 \hbar^2\omega_e'}
\end{equation}
where $\omega_e'=\omega_e [2A_{ab}B/\left(A_{ab}^2+4B^2 \right) ]^{1/2}$ slightly modifies $\omega_e$ 
($\omega_e'\approx 0.9\,\omega_e$).\\
In the limit of high temperature, our model simplifies to 
\begin{equation}
\lim_{T\rightarrow\infty}\, \Tm{63}{c}=2 \,^{63}V^o_{ab}\,\tau_2=\frac{2}{4\hbar^2}\left(A_{ab}^2+4B^2 \right)\tau_2
\end{equation}
and therefore the relation
\begin{equation}
\tau_2=\sqrt{\frac{\pi}{2}}\frac{1}{\omega_e'}
\end{equation}
connects our correlation time $\tau_2$ to the correlation frequency of a magnetic moment in the paramagnetic regime of an insulating antiferromagnet. It must be stressed, however, that in the present work the time $\tau_2$ had been introduced as an independent contribution to $\taueffm(T)$, i.e. 
$\taueffm^{-1}=\tau_1^{-1}+\tau_2^{-1}$, which gave an excellent fit to the relaxation rate data in {\it metallic} \ybco{7} for 100 K $<$ T $<$ 300 K, with $\tau_1$ linear in $T$ and $\tau_2$ constant.

To complement these qualitative findings with a quantitative result we present 
in Fig.~\ref{fig:imailasco15} the data reported by Imai {\it et al.}~\cite{imaisli:93} (full squares) and Ohsugi {\it et al.}~\cite{ohsugi} (empty squares) for the $x$=0.15 samples together with a fit of the data from the former source,
according to (\ref{taumodgap})-(\ref{tausgap}). 
\begin{figure}
\begin{center}
\includegraphics*[width=8cm]{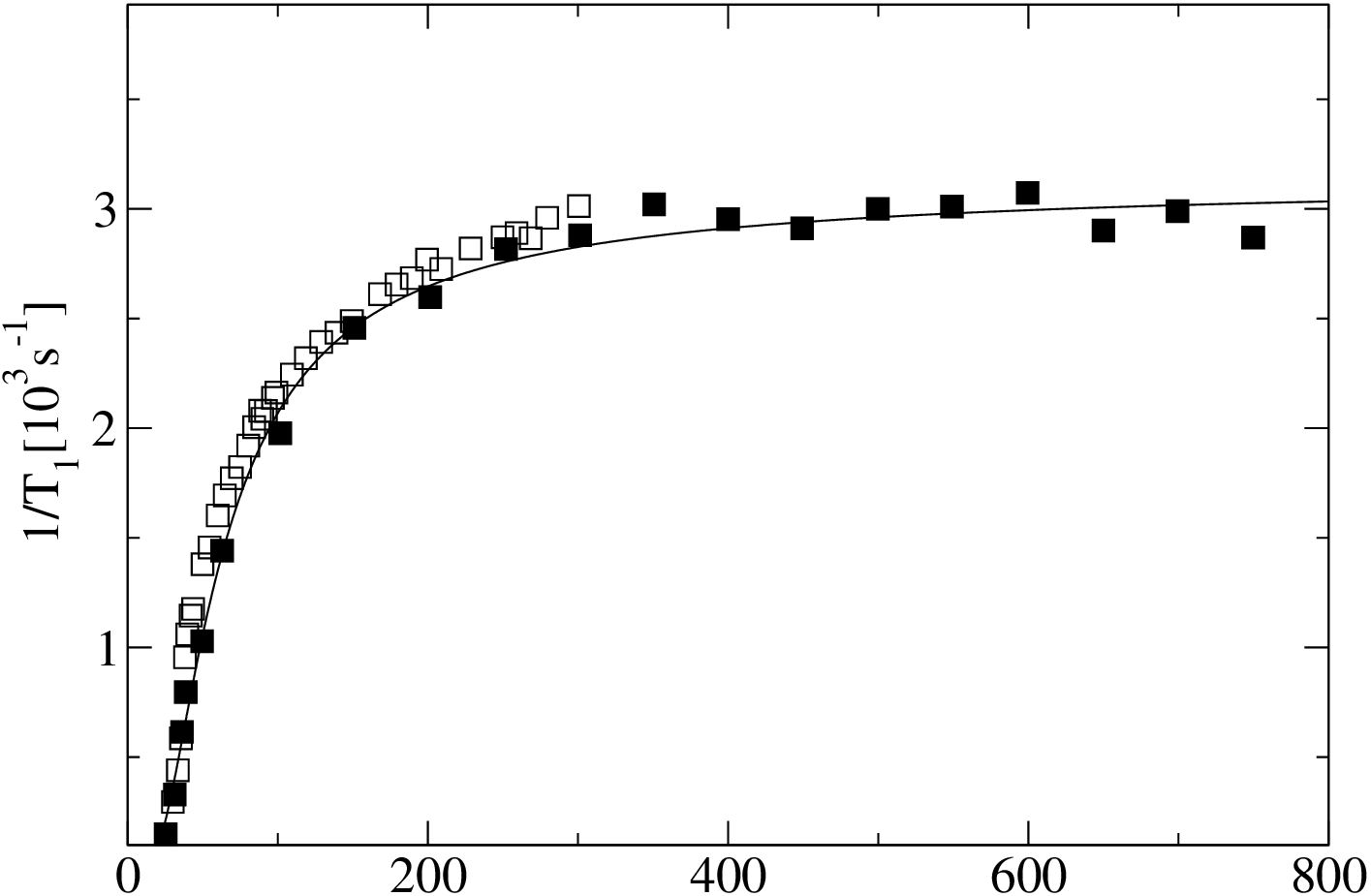}
\caption{\lasco data from Imai {\it et al.}~\cite{imaisli:93} (full squares) and Ohsugi {\it et al.}~\cite{ohsugi} (for $x=0.15$, empty squares). \label{fig:imailasco15}}
\end{center}
\end{figure}
The fit, for which we have assumed a constant value
for $\Vsn{63}{ab}$, gives $\Vsn{63}{ab}a=57.1$ (sK)$^{-1}$, $\Vsn{63}{ab}\tau_2=1574$ s$^{-1}$ and $g=64.1$ K. Adopting the hyperfine interaction energies for \lacuo reported by Haase {\it et al.}~\cite{haase:00}, we find $a=106 \times 10^{-18}$ s/K and $\tau_2=2.92 \times10^{-15}$ s. It is interesting to note that this value of $\tau_2$ is very close to those obtained for \ybco{7} and 
\ybco{6.63}.


\section{Summary and conclusions}\label{sec:sum}
We have compiled a complete set of relaxation rates data for optimally doped \ybco{7} from $T_c$ up to room temperature and extracted $\us{k}{\alpha}$, the contributions to the rates from the fluctuations along the individual crystallographic axes. The result of this simple data transformation is that from this point of view there is no striking difference between the copper and the oxygen relaxations. This suggested the decoupling $\us{k}{\alpha}(T)=\Vs{k}{\alpha}(T)\taueffm(T)$, where the temperature dependence of $\Vs{k}{\alpha}$ is given by the static AFM correlations $\Ks{01}{\alpha}(T)$ for the oxygen and $\Ks{01}{\alpha}(T)$, $\Ks{12}{\alpha}(T)$ and $\Ks{13}{\alpha}(T)$ for the Cu. A numerical fit to the data on $\taueffm$-independent ratios of relaxation rates yielded the correlation lengths in function of the temperature as well as refined hyperfine field constants. The validity of our approach is confirmed by the fact that the seven $\taus{k}{\alpha}(T)$ are well grouped. The average, assimilated to the effective correlation time $\taueffm$, is extremely well modeled by $\taueffm^{-1}(T) = \tau_1^{-1}(T) + \tau_2^{-1}$ with $\tau_2$ constant and $\tau_1(T)$ linear in $T$. This led to the surprising result that in \ybco{7} the ``basic'' relaxation of all three nuclei under consideration is dominated at low temperature by scattering processes of fermionic excitations. The extrapolation of the model predictions to higher temperatures is in very good agreement with the 
measurements. A similar but slightly reduced analysis was conducted on \ybco{6.63}, leading as expected to higher values for the correlations. In the case of \ybcoE we had an even 
more reduced set of data, however some very precise data for Cu are available at high temperature. The analysis of the effective correlation time $\taueffm(T)$ in both underdoped compounds revealed that $\taueffm^{-1}(T) = \tau_1^{-1}(T) + \tau_2^{-1}$ fits the result again very well, provided that $\tau_1(T)$ is modified by a gap function at lower temperature. From the analysis of the \lasco series we could connect $\tau_2$ with relaxation due to AFM spin fluctuations in a paramagnetic state.\\

In conclusion, we have shown that in the model of fluctuating fields the AFM spin-spin correlations $\Ks{01}{\alpha}(T)$, $\Ks{12}{\alpha}(T)$ and 
$\Ks{13}{\alpha}(T)$ determine the degree of coherency. In our fit to the data we found that the in-plane correlations are about $-0.4$ for \ybco{7} and $-0.65$ for \ybcoE at $100$ K. The copper and yttrium relaxation rates depend on all three correlations $\Ks{01}{\alpha}(T)$, $\Ks{12}{\alpha}(T)$ and 
$\Ks{13}{\alpha}(T)$, whereas $\Tm{17}{\alpha}$ involves only $\Ks{01}{\alpha}$. Moreover, the contributions of the three correlations at the copper partially cancel out due to the particular values of the hyperfine field constants. It would be therefore extremely useful to collect more measurements on oxygen, in all crystallographic directions and in all substances.\\
We also note that whereas the particular choice of spatial exponential decay of the correlations fixes the details of the correlation lengths and hyperfine field constants, the same general conclusions of our model would be drawn if another form of spatial decay would be applied. Another note is that the question of the degree of coherency should be addressed at an even lower level than we do, since one should also consider how the various contribution to the {\it on-site} hyperfine fields are added together. For simplicity however we postpone this discussion to a future publication.\\

The essential asset of the model presented in this paper is that a same (fast) fluctuation mechanism of the spin liquid can be identified in the relaxation of all the nuclei (copper, oxygen and yttrium), and this mechanism is characterized by a temperature-dependent effective correlation time $\taueffm(T)$. Moreover, the parameterization ${\taueffm}^{-1}={\tau_1}^{-1}+\tau_2^{-1}$ has a wide range of validity. At low temperature, the effective correlation time $\taueffm$ is just $\tau_1$, which is linear in \ybco{7} but is modified by a gap function in the underdoped compounds. In \ybco{7} the term $\tau_1$ could be linked to the nuclear relaxation by charge carriers in metals. At high temperature, the effective correlation time $\taueffm$ goes over to the constant $\tau_2$, which could be connected to the correlation time in antiferromagnets. This seems to indicate that at high temperature, the nuclei are probably influenced by a system of local moments not very different from the paramagnetic phase of the underdoped parent compounds. Lowering the temperature, a smooth crossover to an itinerant system is 
observed. It looks like these features which have been observed by Imai {\it et al.}\cite{imaisli:93} in \lacuo are also present in the YBaCuO system although the crossover happens at higher temperatures. At low temperature, even in the optimally doped \ybco{7}, a seemingly local pairing of 
moments occurs that is reflected in the Curie-Weiss behavior of $\chir{} (\vec{Q})$. Further analysis, however, is required to get more quantitative results about the doping dependence of all these phenomena. It should be mentioned that also heavy fermions and mixed valent systems exhibit a crossover from localized moments to coherent behavior with decreasing temperature, as has been pointed out recently~\cite{curropines}. \\
We also note that the incommensuration or discommensuration of the AFM ordering observed by neutron scattering measurements does not invalidate the model presented here~\footnote{We recall that the 
essential quantity in the present approach is the nearest neighbor spin correlation
$K_{01}^{\alpha}$ which, for \ybco{7}, changes from $-0.4$ at $T$~=~100~K to $-0.2$ at room
temperatures. Whether these values are due to incommensurability or discommensurability
accompanied with fast fluctuations does not matter. Differences will occur in the
correlations $K_{12}^{\alpha}$ and $K_{13}^{\alpha}$. A detailed investigation
will be published elsewhere.}. Moreover, the model (\ref{tausgap}) and (\ref{taumodgap}) has a wide range of application, extending also to electron-doped superconductors. In particular, the NMR measurements taken by Imai {\it et al.}~\cite{imaisli:edoped} on Sr$_{0.9}$La$_{0.1}$CuO$_2$ can also be fitted within our model. \\

Another important conclusion is that having identified a similar ``basic'' nuclear relaxation rate for all nuclei, the different temperature behavior of the observed rate of the copper and oxygen can be put down to the particular values of the hyperfine field constants. A detailed discussion of the functions $\Vs{k}{\alpha}(T)$ showed the crucial interplay of nearest neighbors and higher AFM correlations and revealed the complications one meets in trying to find Korringa relations in layered cuprates. A case of practical interest is that the on-site and transferred hyperfine energies for Cu are such that $\Vs{63}{ab}(T)$ (which determines $\Tm{63}{c}$) changes only slightly since the contributions from the NN and the third neighbor correlations nearly cancel each other out.\\

We proposed in this paper an alternative way of looking at the spin-relaxation data which we believe could help identifying the underlying physical processes in the cuprates. The analysis of the selected set of data in cuprates shows that the present model reveals the wealth of information that can be obtained from NMR and NQR measurements. While the success of the parameterization $\taueffm^{-1}(T) = \tau_1^{-1}(T) + \tau_2^{-1}$ is astonishingly good and universal, we could only suggest likely explanations about its significance, and a conclusive interpretation is still open to speculations.\\


\begin{acknowledgments}
We would like to thank A. H\"ochner, who determined in the framework of his diploma thesis the hyperfine field constants used as starting values in this work. 
We are particularly grateful to M. Mali and J. Roos for providing much insight in experimental work and for many fruitful discussions. We also appreciated interesting discussions with D. Brinkmann, H. R Ott and T. M. Rice. We thank also our colleagues E. Stoll and S. Renold for their valuable inputs. One of us (PFM) thanks C. P. Slichter for inspiring discussions. This work was supported by the Swiss National Science Foundation.
\end{acknowledgments}


\appendix


\section{Scaled contributions to the relaxation rates}\label{app:Ubars}
The quantities $\us{k}{\alpha}$ extracted from the experiments can be gathered into one plot, each on a different scale. This is equivalent to applying the affine 
transformation 
\begin{equation}
\usbar{k}{\alpha} = \,{^kf}_{\alpha}\, \us{k}{\alpha} + \,{^kg}_{\alpha}
\end{equation}
with constants ${^k f}_{\alpha}$ and ${^k g}_{\alpha}$. The result for \ybco{7} is shown in Fig. \ref{fig:Usrenorm7}, where the vertical axis is arbitrary.
\begin{figure}
\begin{center}
\subfigure[\ybco{7}]{\label{fig:Usrenorm7}
\includegraphics*[width=8cm]{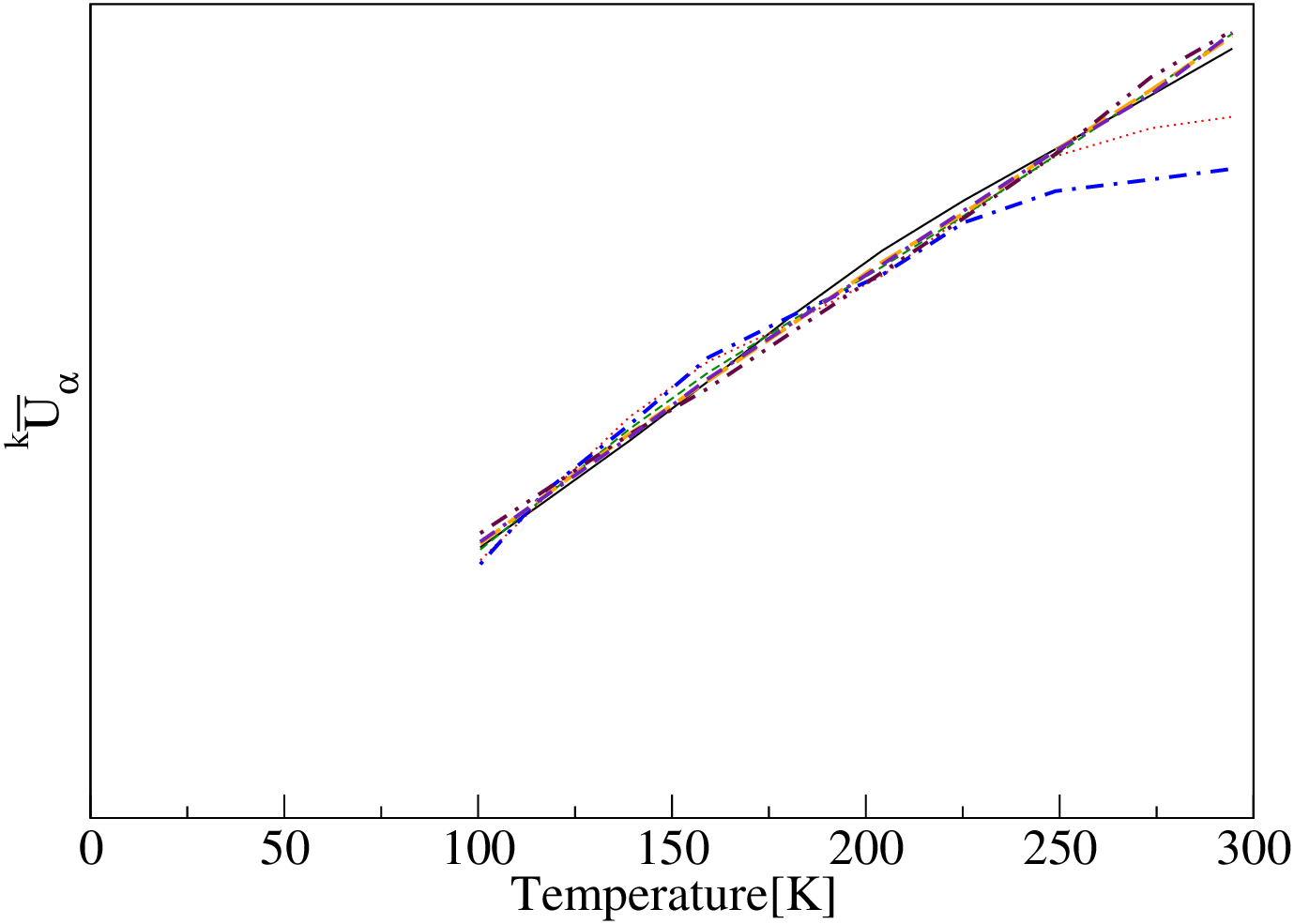}}
\subfigure[\ybco{6.63}]{\label{fig:Usrenorm663}
\includegraphics*[width=8cm]{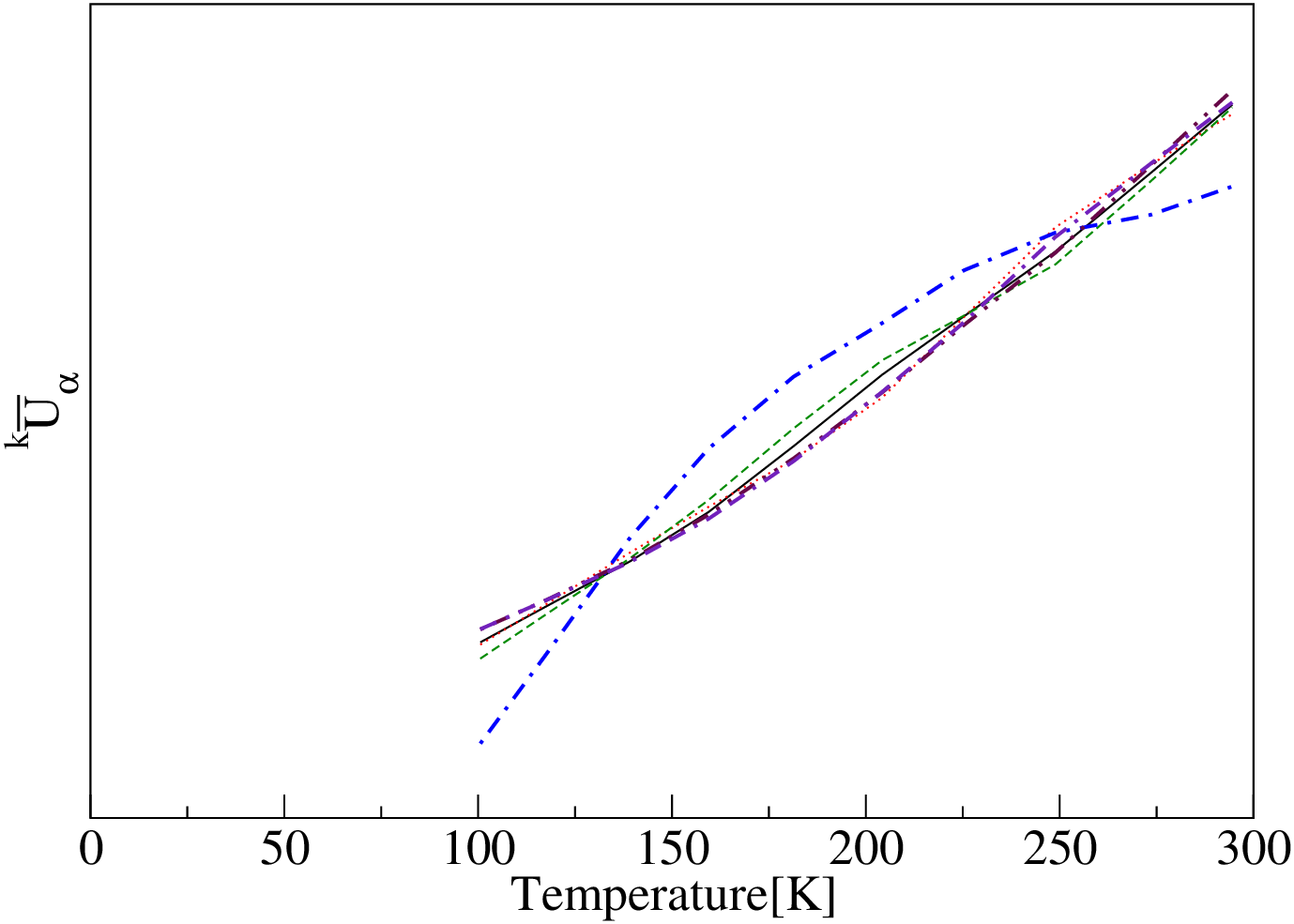}}
\caption{The $\us{k}{\alpha}$ 
after affine transformations of the 
data. $\usbar{17}{a}$ (thin full line), 
 $\usbar{17}{b}$ (thin dotted), 
 $\usbar{17}{c}$ (thin dashed), 
 $\usbar{63}{ab}$ (thick dash-dotted), 
 $\usbar{63}{c}$ (thick dashed), 
 $\usbar{89}{ab}$ (thick dash-dot-dot-dashed), 
 $\usbar{89}{c}$ (thick dot-dash-dash-dotted). \label{fig:Usrenorm}}
\end{center}
\end{figure}
We note that the $\us{k}{\alpha}$ in \ybco{7} have the same temperature dependence for most of the temperature range. The larger deviations occur at higher temperature for $\us{63}{ab}$ (thick dash-dots) and $\us{17}{b}$ (thin dots). The same operations can be done for \ybco{6.63}, Fig. \ref{fig:Usrenorm663}. Again all $\us{k}{\alpha}$ fall on the same curve, apart $\us{63}{ab}$ which deviates significantly.\\
A similar picture is drawn upon dividing the rates $\us{k}{\alpha}(T)$ by the constants $\Vsn{k}{\alpha}$ that are calculated from (\ref{V0}) and whose values are gathered in Table \ref{tab:V0}: $\ustil{k}{\alpha}(T):=\,\us{k}{\alpha}(T)/\Vsn{k}{\alpha}$. The scaled relaxation rates $\ustil{k}{\alpha}(T)$ are shown in Fig. \ref{fig:Util} for \ybco{7} 
and \ybco{6.63}.
\begin{figure}
\begin{center}
\subfigure[\ybco{7}]{\label{fig:Util7}
\includegraphics*[width=8cm]{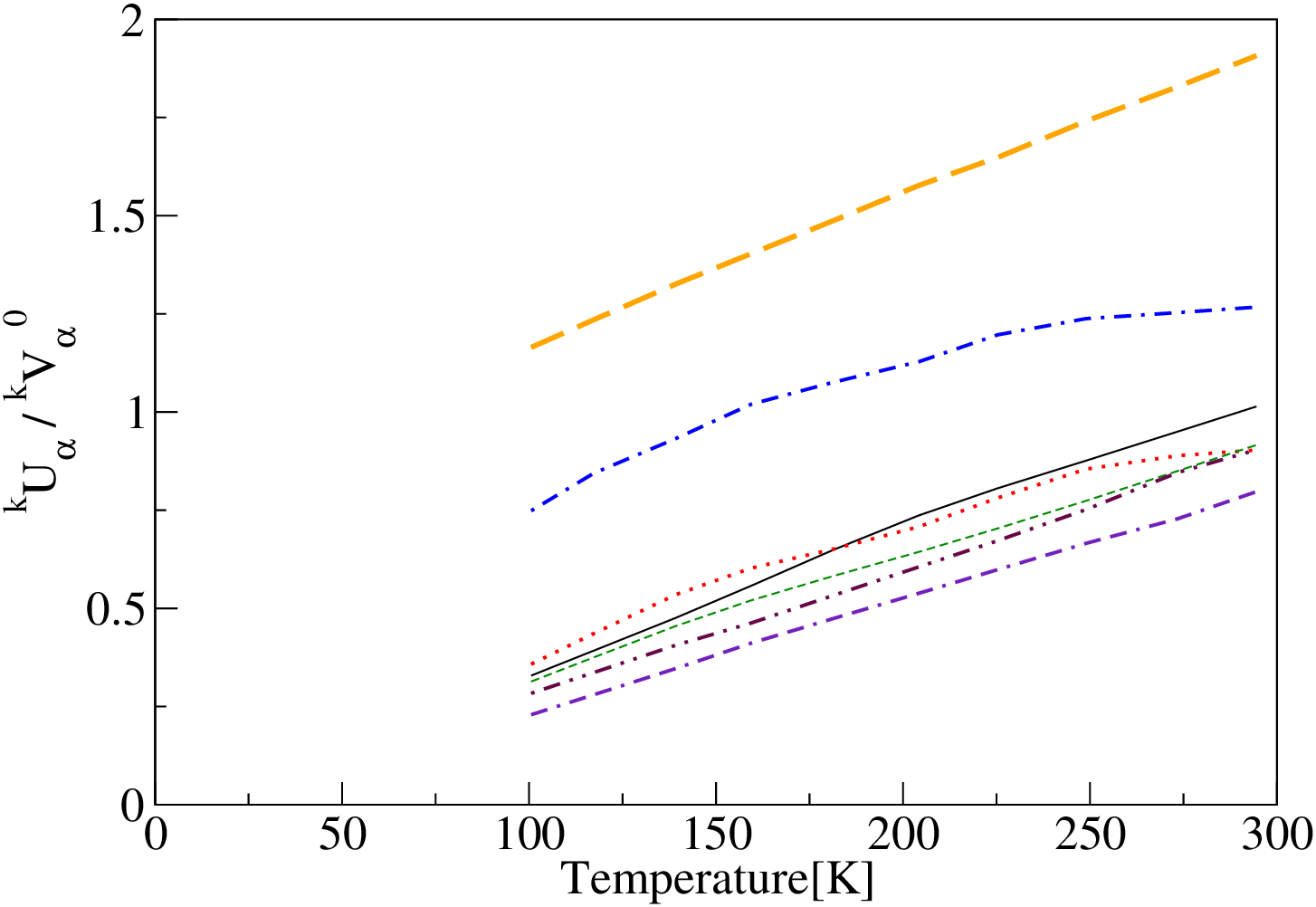}}
\subfigure[\ybco{6.63}]{\label{fig:Util6}
\includegraphics*[width=8cm]{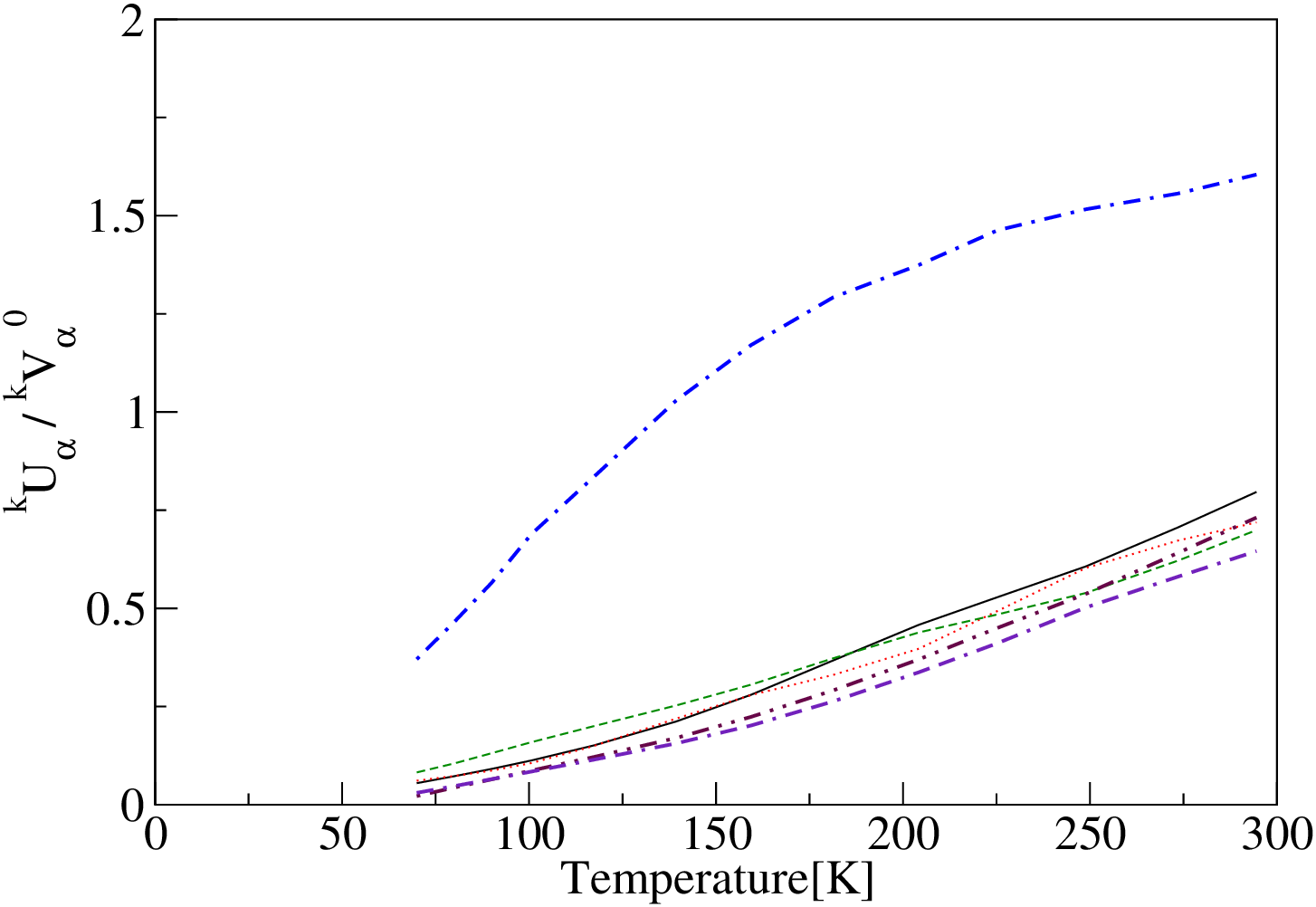}}
\caption{$\ustil{k}{\alpha}(T)=\,\us{k}{\alpha}(T)/\Vsn{k}{\alpha}$. 
$\ustil{17}{a}$ (thin full line), 
 $\ustil{17}{b}$ (thin dotted), 
 $\ustil{17}{c}$ (thin dashed), 
 $\ustil{63}{ab}$ (thick dash-dotted), 
 $\ustil{63}{c}$ (thick dashed), 
 $\ustil{89}{ab}$ (thick dash-dot-dot-dashed), 
 $\ustil{89}{c}$ (thick dot-dash-dash-dotted). \label{fig:Util}}
\end{center}
\end{figure}

\section{Detailed Investigation of $^kV_{\alpha}(T)$ and Korringa-like behavior}\label{app:V}
We plotted in Fig. \ref{fig:V63dec_7} the individual contributions 
to $\Vs{63}{\alpha}(T)$, which are defined as
\begin{eqnarray}
\Q{63}{\alpha}{1}(T) &=& \frac{1}{4\hbar^2}[8 A_{\alpha} B K^{\alpha}_{01} (T) ]  \nonumber \\
\Q{63}{\alpha}{2}(T) &=& \frac{1}{4\hbar^2}[8 B^2 
K^{\alpha}_{12}(T) ] \nonumber \\
\Q{63}{\alpha}{3}(T) &=& \frac{1}{4\hbar^2}[4 B^2 K^{\alpha}_{13}(T) ]
\end{eqnarray}
$\Q{63}{\alpha}{1}$, $\Q{63}{\alpha}{2}$ and $\Q{63}{\alpha}{3}$ are the terms proportional respectively to the AFM spin correlation functions $K^{\alpha}_{01}$, $K^{\alpha}_{12}$ and $K^{\alpha}_{13}$. We also have the constant term $\Q{63}{\alpha}{0}$, the temperature independent contribution given in Eq. (\ref{V063}) and Table \ref{tab:V0}.
\begin{figure}
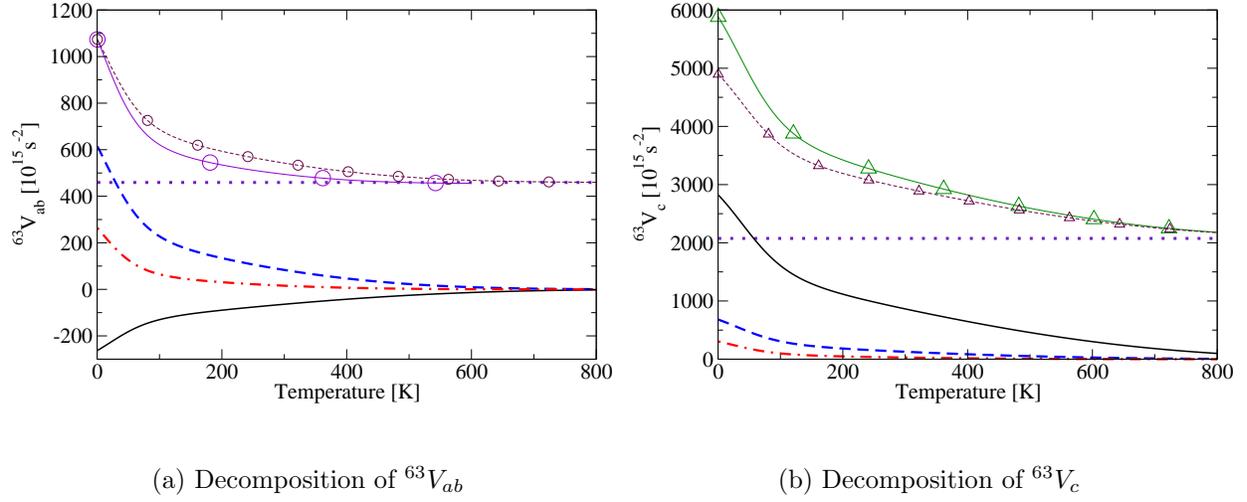

\begin{center}
\subfigure[Decomposition of $^{63}V_{ab}$]{\label{fig:V63dec_7ab}
\includegraphics*[width=8cm]{UtauCuT_4exABd.eps}}
\subfigure[Decomposition of $^{63}V_{c}$]{\label{fig:V63dec_7c}
\includegraphics*[width=8cm]{UtauCuT_4exCd.eps}}
\caption{\ybco{7}. Full $^{63}V_{\alpha}$ (large symbols), $\Q{63}{\alpha}{0}$ (dotted line), $\Q{63}{\alpha}{1}$ (full line), $\Q{63}{\alpha}{2}$ (dashed line), $\Q{63}{\alpha}{3}$ (dash-dotted line). $\Q{63}{ab}{0}+\,\Q{63}{ab}{2}$ (small circles), $\Q{63}{c}{0}+\,\Q{63}{c}{1}$ (small triangles). \label{fig:V63dec_7}}
\end{center}
\end{figure}
Since $A_{ab} > 0$, $A_c < 0$, $\Ks{01}{\alpha} \leq 0$ and $\Ks{12}{\alpha},\Ks{13}{\alpha} \geq 0$, all individual contributions $\Q{63}{\alpha}{i}$ are positive except 
$\Q{63}{ab}{1}$. We see in Fig. \ref{fig:V63dec_7ab} that $\Vs{63}{ab}$ (large circles) is made up almost entirely of $\Q{63}{ab}{0}+\Q{63}{ab}{2}$ (small circles), whereas $\Vs{63}{c}$ in Fig. \ref{fig:V63dec_7c} (large triangles) is dominated by $\Q{63}{c}{0}+\Q{63}{c}{1}$ (small triangles). In other words, in the ab-direction $\Q{63}{ab}{1}$ and $\Q{63}{ab}{3}$, having opposite signs, almost cancel each other out. For example near $T_c$ where the correlations are high,
$|\Q{63}{ab}{1}/ \Q{63}{ab}{3}|=2 \frac{A_{ab}}{B} |\Ks{01}{ab}|/|\Ks{01}{ab}|^2 \approx 1$. In the c-direction all contributions have the same sign but due to the high value of $|A_c|$ the lower order correlation plays the major role. Note that the same qualitative behavior is expected irrespective of the choice of correlation dependence 
(Eq. \ref{Ks}).\\

The seemingly Korringa-like behavior of $\TTm{17}{\alpha}$ is investigated in 
Fig. \ref{fig:tauzerl_7ab}.
\begin{figure}
\begin{center}
\includegraphics*[width=8cm]{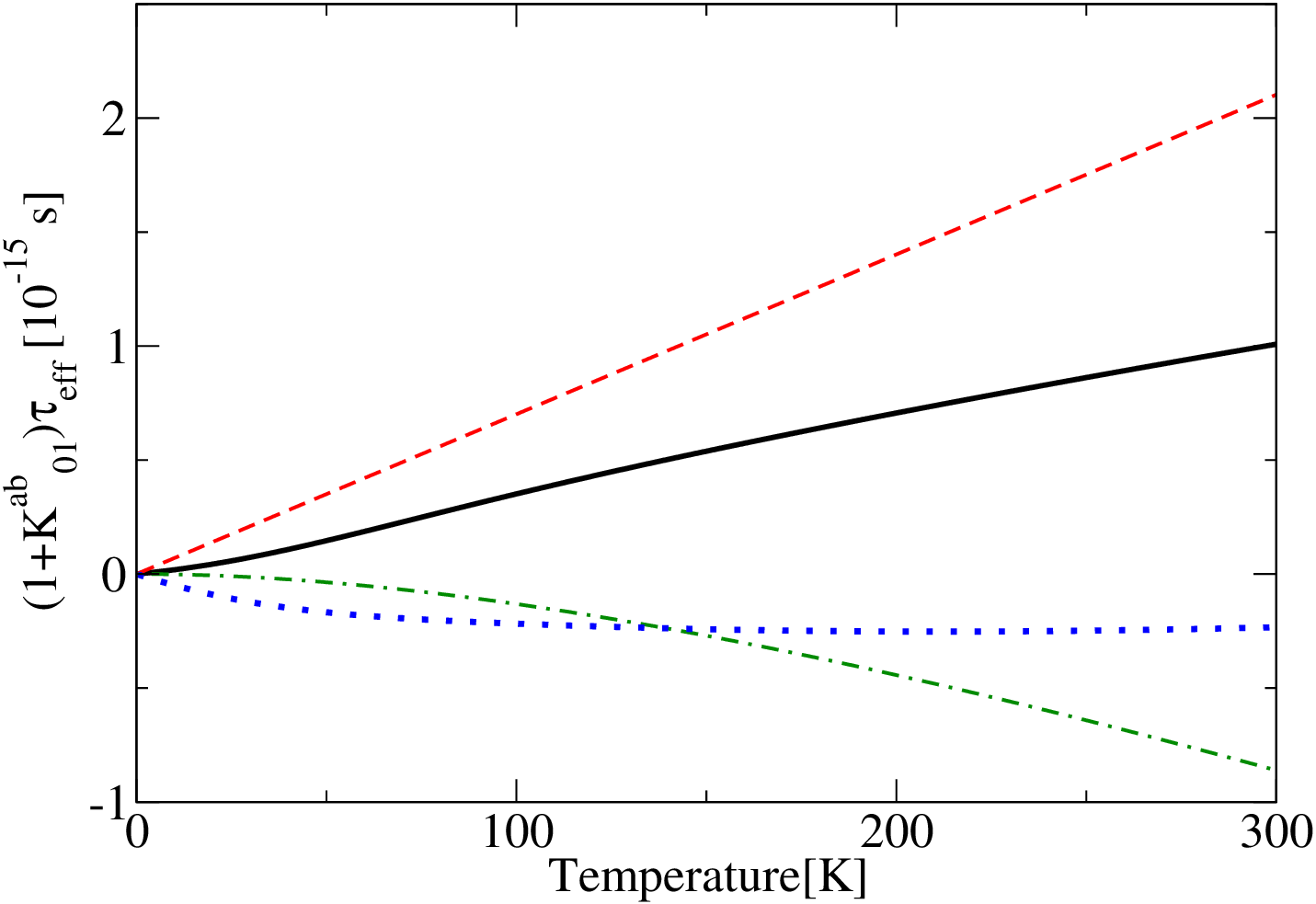}
\caption{\ybco{7}. Decomposition of 
$(1+K^{ab}_{01})\tau_{\mathit{eff}}$ (full line): $\tau_1$ (dashed line), $\taueffm-\tau_1$ (dash-dotted line) and $K^{ab}_{01}\taueffm$ (dotted line). \label{fig:tauzerl_7ab}}
\end{center}
\end{figure}
There we decomposed $\Tm{17}{c}\propto (1-|\Ks{01}{ab}|)\taueffm$ (full line) 
into 
\begin{equation}
(1-|\Ks{01}{ab}|)\taueffm=\tau_1+(\taueffm-\tau_1)-\taueffm |\Ks{01}{ab}|
\end{equation}
From the figure \ref{fig:tauzerl_7ab} we see that the contribution $\taueffm |\Ks{01}{ab}|$ (dotted line) varies little over a large range of temperature. The dashed line is the truly linear contribution $\tau_1$. The actual temperature dependence of $\TTm{17}{\alpha}$ is thus given by this linear contribution, minus that of $(\taueffm-\tau_1)$ (dash-dotted line). The total result has however the appearance of linearity.


\bibliography{biblio}

\begin{thebibliography}{47}
\expandafter\ifx\csname natexlab\endcsname\relax\def\natexlab#1{#1}\fi
\expandafter\ifx\csname bibnamefont\endcsname\relax
  \def\bibnamefont#1{#1}\fi
\expandafter\ifx\csname bibfnamefont\endcsname\relax
  \def\bibfnamefont#1{#1}\fi
\expandafter\ifx\csname citenamefont\endcsname\relax
  \def\citenamefont#1{#1}\fi
\expandafter\ifx\csname url\endcsname\relax
  \def\url#1{\texttt{#1}}\fi
\expandafter\ifx\csname urlprefix\endcsname\relax\def\urlprefix{URL }\fi
\providecommand{\bibinfo}[2]{#2}
\providecommand{\eprint}[2][]{\url{#2}}

\bibitem[{\citenamefont{Brinkmann and Mali}(1994)}]{book:mali}
\bibinfo{author}{\bibfnamefont{D.}~\bibnamefont{Brinkmann}} \bibnamefont{and}
  \bibinfo{author}{\bibfnamefont{M.}~\bibnamefont{Mali}},
  \emph{\bibinfo{title}{NMR Basic Principles and Progress}},
  vol.~\bibinfo{volume}{31} (\bibinfo{publisher}{Springer},
  \bibinfo{address}{Heidelberg}, \bibinfo{year}{1994}).

\bibitem[{\citenamefont{Slichter}(1993)}]{rev:slichter}
\bibinfo{author}{\bibfnamefont{C.~P.} \bibnamefont{Slichter}}, in
  \emph{\bibinfo{booktitle}{Strongly Correlated Electronic Materials, The Los
  Alamos Symposium}}, edited by \bibinfo{editor}{\bibfnamefont{K.~S.}
  \bibnamefont{Bedell}},
  \bibinfo{editor}{\bibfnamefont{Z.}~\bibnamefont{Wang}},
  \bibinfo{editor}{\bibfnamefont{D.}~\bibnamefont{Meltzer}},
  \bibinfo{editor}{\bibfnamefont{A.~V.} \bibnamefont{Balatsky}},
  \bibnamefont{and} \bibinfo{editor}{\bibfnamefont{E.}~\bibnamefont{Abrahams}}
  (\bibinfo{publisher}{Addison-Wesley}, \bibinfo{year}{1993}), p.
  \bibinfo{pages}{427}.

\bibitem[{\citenamefont{Berthier et~al.}(1996)\citenamefont{Berthier, Julien,
  Horvatic, and Berthier}}]{berthier}
\bibinfo{author}{\bibfnamefont{C.}~\bibnamefont{Berthier}},
  \bibinfo{author}{\bibfnamefont{M.~H.} \bibnamefont{Julien}},
  \bibinfo{author}{\bibfnamefont{M.}~\bibnamefont{Horvatic}}, \bibnamefont{and}
  \bibinfo{author}{\bibfnamefont{Y.}~\bibnamefont{Berthier}},
  \bibinfo{journal}{J.Phys. I France} \textbf{\bibinfo{volume}{6}},
  \bibinfo{pages}{2205} (\bibinfo{year}{1996}).

\bibitem[{\citenamefont{Rigamonti et~al.}(1998)\citenamefont{Rigamonti, Borsa,
  and Caretta}}]{rigamonti}
\bibinfo{author}{\bibfnamefont{A.}~\bibnamefont{Rigamonti}},
  \bibinfo{author}{\bibfnamefont{F.}~\bibnamefont{Borsa}}, \bibnamefont{and}
  \bibinfo{author}{\bibfnamefont{P.}~\bibnamefont{Caretta}},
  \bibinfo{journal}{Rep. Prog. Phys.} \textbf{\bibinfo{volume}{61}},
  \bibinfo{pages}{1367} (\bibinfo{year}{1998}).

\bibitem[{\citenamefont{Warren et~al.}(1987)\citenamefont{Warren, Walstedt,
  Brennert, Espinosa, and Remeika}}]{warren:87}
\bibinfo{author}{\bibfnamefont{W.~W.} \bibnamefont{Warren}},
  \bibinfo{author}{\bibfnamefont{R.~E.} \bibnamefont{Walstedt}},
  \bibinfo{author}{\bibfnamefont{G.~F.} \bibnamefont{Brennert}},
  \bibinfo{author}{\bibfnamefont{G.~P.} \bibnamefont{Espinosa}},
  \bibnamefont{and} \bibinfo{author}{\bibfnamefont{J.~P.}
  \bibnamefont{Remeika}}, \bibinfo{journal}{Phys. Rev. Lett.}
  \textbf{\bibinfo{volume}{59}}, \bibinfo{pages}{1860} (\bibinfo{year}{1987}).

\bibitem[{\citenamefont{Imai et~al.}(1988{\natexlab{a}})\citenamefont{Imai,
  Shimizu, Tsuda, Yasuoka, Takabatake, and Nakazawa}}]{imai1:88}
\bibinfo{author}{\bibfnamefont{T.}~\bibnamefont{Imai}},
  \bibinfo{author}{\bibfnamefont{T.}~\bibnamefont{Shimizu}},
  \bibinfo{author}{\bibfnamefont{T.}~\bibnamefont{Tsuda}},
  \bibinfo{author}{\bibfnamefont{H.}~\bibnamefont{Yasuoka}},
  \bibinfo{author}{\bibfnamefont{T.}~\bibnamefont{Takabatake}},
  \bibnamefont{and} \bibinfo{author}{\bibfnamefont{Y.}~\bibnamefont{Nakazawa}},
  \bibinfo{journal}{J. Phys. Soc. Jpn} \textbf{\bibinfo{volume}{57}},
  \bibinfo{pages}{1771} (\bibinfo{year}{1988}{\natexlab{a}}).

\bibitem[{\citenamefont{Imai et~al.}(1988{\natexlab{b}})\citenamefont{Imai,
  Shimizu, Yasuoka, Ueda, and Kosuge}}]{imai2:88}
\bibinfo{author}{\bibfnamefont{T.}~\bibnamefont{Imai}},
  \bibinfo{author}{\bibfnamefont{T.}~\bibnamefont{Shimizu}},
  \bibinfo{author}{\bibfnamefont{H.}~\bibnamefont{Yasuoka}},
  \bibinfo{author}{\bibfnamefont{Y.}~\bibnamefont{Ueda}}, \bibnamefont{and}
  \bibinfo{author}{\bibfnamefont{K.}~\bibnamefont{Kosuge}},
  \bibinfo{journal}{J. Phys. Soc. Jpn} \textbf{\bibinfo{volume}{57}},
  \bibinfo{pages}{2280} (\bibinfo{year}{1988}{\natexlab{b}}).

\bibitem[{\citenamefont{Hammel et~al.}(1989)\citenamefont{Hammel, Takigawa,
  Heffner, Fisk, and Ott}}]{hammel:89}
\bibinfo{author}{\bibfnamefont{P.~C.} \bibnamefont{Hammel}},
  \bibinfo{author}{\bibfnamefont{M.}~\bibnamefont{Takigawa}},
  \bibinfo{author}{\bibfnamefont{R.~H.} \bibnamefont{Heffner}},
  \bibinfo{author}{\bibfnamefont{Z.}~\bibnamefont{Fisk}}, \bibnamefont{and}
  \bibinfo{author}{\bibfnamefont{K.~C.} \bibnamefont{Ott}},
  \bibinfo{journal}{Phys. Rev. Lett.} \textbf{\bibinfo{volume}{63}},
  \bibinfo{pages}{1992} (\bibinfo{year}{1989}).

\bibitem[{\citenamefont{Warren et~al.}(1989)\citenamefont{Warren, Walstedt,
  Brennert, Cava, Tycko, Bell, and Dabbagh}}]{warren:89}
\bibinfo{author}{\bibfnamefont{W.~W.} \bibnamefont{Warren}},
  \bibinfo{author}{\bibfnamefont{R.~E.} \bibnamefont{Walstedt}},
  \bibinfo{author}{\bibfnamefont{G.~F.} \bibnamefont{Brennert}},
  \bibinfo{author}{\bibfnamefont{R.~J.} \bibnamefont{Cava}},
  \bibinfo{author}{\bibfnamefont{R.}~\bibnamefont{Tycko}},
  \bibinfo{author}{\bibfnamefont{R.~F.} \bibnamefont{Bell}}, \bibnamefont{and}
  \bibinfo{author}{\bibfnamefont{G.}~\bibnamefont{Dabbagh}},
  \bibinfo{journal}{Phys. Rev. Lett.} \textbf{\bibinfo{volume}{62}},
  \bibinfo{pages}{1193} (\bibinfo{year}{1989}).

\bibitem[{\citenamefont{Walstedt et~al.}(1990)\citenamefont{Walstedt, Warren,
  Bell, Cava, Espinosa, Schneemeyer, and Waszczak}}]{wals:90}
\bibinfo{author}{\bibfnamefont{R.~E.} \bibnamefont{Walstedt}},
  \bibinfo{author}{\bibfnamefont{W.~W.} \bibnamefont{Warren}},
  \bibinfo{author}{\bibfnamefont{R.~F.} \bibnamefont{Bell}},
  \bibinfo{author}{\bibfnamefont{R.~J.} \bibnamefont{Cava}},
  \bibinfo{author}{\bibfnamefont{G.~P.} \bibnamefont{Espinosa}},
  \bibinfo{author}{\bibfnamefont{L.~F.} \bibnamefont{Schneemeyer}},
  \bibnamefont{and} \bibinfo{author}{\bibfnamefont{J.~V.}
  \bibnamefont{Waszczak}}, \bibinfo{journal}{Phys. Rev. B}
  \textbf{\bibinfo{volume}{41}}, \bibinfo{pages}{R9574} (\bibinfo{year}{1990}).

\bibitem[{\citenamefont{Imai et~al.}(1990)\citenamefont{Imai, Yoshimura,
  Uemura, Yosuoka, and Kosuge}}]{imai:90}
\bibinfo{author}{\bibfnamefont{T.}~\bibnamefont{Imai}},
  \bibinfo{author}{\bibfnamefont{K.}~\bibnamefont{Yoshimura}},
  \bibinfo{author}{\bibfnamefont{T.}~\bibnamefont{Uemura}},
  \bibinfo{author}{\bibfnamefont{H.}~\bibnamefont{Yosuoka}}, \bibnamefont{and}
  \bibinfo{author}{\bibfnamefont{K.}~\bibnamefont{Kosuge}},
  \bibinfo{journal}{J. Phys. Soc. Jpn} \textbf{\bibinfo{volume}{59}},
  \bibinfo{pages}{3846} (\bibinfo{year}{1990}).

\bibitem[{\citenamefont{Takigawa et~al.}(1991)\citenamefont{Takigawa, Reyes,
  Hammel, Thompson, Heffner, Fisk, and Ott}}]{taki:91}
\bibinfo{author}{\bibfnamefont{M.}~\bibnamefont{Takigawa}},
  \bibinfo{author}{\bibfnamefont{A.~P.} \bibnamefont{Reyes}},
  \bibinfo{author}{\bibfnamefont{P.~C.} \bibnamefont{Hammel}},
  \bibinfo{author}{\bibfnamefont{J.~D.} \bibnamefont{Thompson}},
  \bibinfo{author}{\bibfnamefont{R.~H.} \bibnamefont{Heffner}},
  \bibinfo{author}{\bibfnamefont{Z.}~\bibnamefont{Fisk}}, \bibnamefont{and}
  \bibinfo{author}{\bibfnamefont{K.~C.} \bibnamefont{Ott}},
  \bibinfo{journal}{Phys. Rev. B} \textbf{\bibinfo{volume}{43}},
  \bibinfo{pages}{247} (\bibinfo{year}{1991}).

\bibitem[{\citenamefont{Walstedt et~al.}(1994)\citenamefont{Walstedt, Shastry,
  and Cheong}}]{wals:94}
\bibinfo{author}{\bibfnamefont{R.~E.} \bibnamefont{Walstedt}},
  \bibinfo{author}{\bibfnamefont{B.~S.} \bibnamefont{Shastry}},
  \bibnamefont{and} \bibinfo{author}{\bibfnamefont{S.~W.}
  \bibnamefont{Cheong}}, \bibinfo{journal}{Phys. Rev. Lett.}
  \textbf{\bibinfo{volume}{72}}, \bibinfo{pages}{3610} (\bibinfo{year}{1994}).

\bibitem[{\citenamefont{Mila and Rice}(1989)}]{milarice}
\bibinfo{author}{\bibfnamefont{F.}~\bibnamefont{Mila}} \bibnamefont{and}
  \bibinfo{author}{\bibfnamefont{T.~M.} \bibnamefont{Rice}},
  \bibinfo{journal}{Physica C} \textbf{\bibinfo{volume}{157}},
  \bibinfo{pages}{561} (\bibinfo{year}{1989}).

\bibitem[{\citenamefont{Monien et~al.}(1990)\citenamefont{Monien, Pines, and
  Slichter}}]{mps}
\bibinfo{author}{\bibfnamefont{H.}~\bibnamefont{Monien}},
  \bibinfo{author}{\bibfnamefont{D.}~\bibnamefont{Pines}}, \bibnamefont{and}
  \bibinfo{author}{\bibfnamefont{C.~P.} \bibnamefont{Slichter}},
  \bibinfo{journal}{Phys. Rev. B} \textbf{\bibinfo{volume}{41}},
  \bibinfo{pages}{11120} (\bibinfo{year}{1990}).

\bibitem[{\citenamefont{Millis et~al.}(1990)\citenamefont{Millis, Monien, and
  Pines}}]{mmp}
\bibinfo{author}{\bibfnamefont{A.~J.} \bibnamefont{Millis}},
  \bibinfo{author}{\bibfnamefont{H.}~\bibnamefont{Monien}}, \bibnamefont{and}
  \bibinfo{author}{\bibfnamefont{D.}~\bibnamefont{Pines}},
  \bibinfo{journal}{Phys. Rev. B} \textbf{\bibinfo{volume}{42}},
  \bibinfo{pages}{167} (\bibinfo{year}{1990}).

\bibitem[{\citenamefont{Zha et~al.}(1996)\citenamefont{Zha, Barzykin, and
  Pines}}]{zha}
\bibinfo{author}{\bibfnamefont{Y.}~\bibnamefont{Zha}},
  \bibinfo{author}{\bibfnamefont{V.}~\bibnamefont{Barzykin}}, \bibnamefont{and}
  \bibinfo{author}{\bibfnamefont{D.}~\bibnamefont{Pines}},
  \bibinfo{journal}{Phys. Rev. B} \textbf{\bibinfo{volume}{54}},
  \bibinfo{pages}{7561} (\bibinfo{year}{1996}).

\bibitem[{\citenamefont{Slichter}(1996)}]{book:slichter}
\bibinfo{author}{\bibfnamefont{C.~P.} \bibnamefont{Slichter}},
  \emph{\bibinfo{title}{Principles of magnetic resonance}}
  (\bibinfo{publisher}{Sprin\-ger, Berlin}, \bibinfo{year}{1996}).

\bibitem[{\citenamefont{Pennington et~al.}(1989)\citenamefont{Pennington,
  Durand, Slichter, Rice, Bukowski, and Ginsberg}}]{penn}
\bibinfo{author}{\bibfnamefont{C.~H.} \bibnamefont{Pennington}},
  \bibinfo{author}{\bibfnamefont{D.~J.} \bibnamefont{Durand}},
  \bibinfo{author}{\bibfnamefont{C.~P.} \bibnamefont{Slichter}},
  \bibinfo{author}{\bibfnamefont{J.~P.} \bibnamefont{Rice}},
  \bibinfo{author}{\bibfnamefont{E.~D.} \bibnamefont{Bukowski}},
  \bibnamefont{and} \bibinfo{author}{\bibfnamefont{D.~M.}
  \bibnamefont{Ginsberg}}, \bibinfo{journal}{Phys. Rev. B}
  \textbf{\bibinfo{volume}{39}}, \bibinfo{pages}{R2902} (\bibinfo{year}{1989}).

\bibitem[{\citenamefont{Shastry}(1989)}]{shastry}
\bibinfo{author}{\bibfnamefont{B.~S.} \bibnamefont{Shastry}},
  \bibinfo{journal}{Phys. Rev. Lett.} \textbf{\bibinfo{volume}{63}},
  \bibinfo{pages}{1288} (\bibinfo{year}{1989}).

\bibitem[{\citenamefont{H\"usser et~al.}(2000)\citenamefont{H\"usser, Suter,
  Stoll, and Meier}}]{huesser}
\bibinfo{author}{\bibfnamefont{P.}~\bibnamefont{H\"usser}},
  \bibinfo{author}{\bibfnamefont{H.~U.} \bibnamefont{Suter}},
  \bibinfo{author}{\bibfnamefont{E.~P.} \bibnamefont{Stoll}}, \bibnamefont{and}
  \bibinfo{author}{\bibfnamefont{P.~F.} \bibnamefont{Meier}},
  \bibinfo{journal}{Phys. Rev. B} \textbf{\bibinfo{volume}{61}},
  \bibinfo{pages}{1567} (\bibinfo{year}{2000}).

\bibitem[{\citenamefont{Renold et~al.}(2001)\citenamefont{Renold, Plibersek,
  Stoll, Claxton, and Meier}}]{renold}
\bibinfo{author}{\bibfnamefont{S.}~\bibnamefont{Renold}},
  \bibinfo{author}{\bibfnamefont{S.}~\bibnamefont{Plibersek}},
  \bibinfo{author}{\bibfnamefont{E.~P.} \bibnamefont{Stoll}},
  \bibinfo{author}{\bibfnamefont{T.~A.} \bibnamefont{Claxton}},
  \bibnamefont{and} \bibinfo{author}{\bibfnamefont{P.~F.} \bibnamefont{Meier}},
  \bibinfo{journal}{Eur. Phys. J. B} \textbf{\bibinfo{volume}{23}},
  \bibinfo{pages}{3} (\bibinfo{year}{2001}).

\bibitem[{\citenamefont{Suter et~al.}(1997)\citenamefont{Suter, Mali, Roos,
  Brinkmann, Karpinski, and Kaldis}}]{suter}
\bibinfo{author}{\bibfnamefont{A.}~\bibnamefont{Suter}},
  \bibinfo{author}{\bibfnamefont{M.}~\bibnamefont{Mali}},
  \bibinfo{author}{\bibfnamefont{J.}~\bibnamefont{Roos}},
  \bibinfo{author}{\bibfnamefont{D.}~\bibnamefont{Brinkmann}},
  \bibinfo{author}{\bibfnamefont{J.}~\bibnamefont{Karpinski}},
  \bibnamefont{and} \bibinfo{author}{\bibfnamefont{E.}~\bibnamefont{Kaldis}},
  \bibinfo{journal}{Phys. Rev. B} \textbf{\bibinfo{volume}{56}},
  \bibinfo{pages}{5542} (\bibinfo{year}{1997}).

\bibitem[{\citenamefont{Meier}(2001)}]{hono}
\bibinfo{author}{\bibfnamefont{P.~F.} \bibnamefont{Meier}},
  \bibinfo{journal}{Physica C} \textbf{\bibinfo{volume}{364-365}},
  \bibinfo{pages}{411} (\bibinfo{year}{2001}).

\bibitem[{\citenamefont{Walstedt et~al.}(1988)\citenamefont{Walstedt, Warren,
  Bell, Brennert, Espinosa, Cava, Schneemeyer, and Waszczak}}]{wals:88}
\bibinfo{author}{\bibfnamefont{R.~E.} \bibnamefont{Walstedt}},
  \bibinfo{author}{\bibfnamefont{W.~W.} \bibnamefont{Warren}},
  \bibinfo{author}{\bibfnamefont{R.~F.} \bibnamefont{Bell}},
  \bibinfo{author}{\bibfnamefont{G.~F.} \bibnamefont{Brennert}},
  \bibinfo{author}{\bibfnamefont{G.~P.} \bibnamefont{Espinosa}},
  \bibinfo{author}{\bibfnamefont{R.~J.} \bibnamefont{Cava}},
  \bibinfo{author}{\bibfnamefont{L.~F.} \bibnamefont{Schneemeyer}},
  \bibnamefont{and} \bibinfo{author}{\bibfnamefont{J.~V.}
  \bibnamefont{Waszczak}}, \bibinfo{journal}{Phys. Rev. B}
  \textbf{\bibinfo{volume}{38}}, \bibinfo{pages}{R9299} (\bibinfo{year}{1988}).

\bibitem[{\citenamefont{Martindale et~al.}(1998)\citenamefont{Martindale,
  Hammel, Hults, and Smith}}]{martind:98}
\bibinfo{author}{\bibfnamefont{J.~A.} \bibnamefont{Martindale}},
  \bibinfo{author}{\bibfnamefont{P.~C.} \bibnamefont{Hammel}},
  \bibinfo{author}{\bibfnamefont{W.~L.} \bibnamefont{Hults}}, \bibnamefont{and}
  \bibinfo{author}{\bibfnamefont{J.~L.} \bibnamefont{Smith}},
  \bibinfo{journal}{Phys. Rev. B} \textbf{\bibinfo{volume}{57}},
  \bibinfo{pages}{11769} (\bibinfo{year}{1998}).

\bibitem[{\citenamefont{Takigawa et~al.}(1993)\citenamefont{Takigawa, Hults,
  and Smith}}]{taki:93}
\bibinfo{author}{\bibfnamefont{M.}~\bibnamefont{Takigawa}},
  \bibinfo{author}{\bibfnamefont{W.~L.} \bibnamefont{Hults}}, \bibnamefont{and}
  \bibinfo{author}{\bibfnamefont{J.~L.} \bibnamefont{Smith}},
  \bibinfo{journal}{Phys. Rev. Lett.} \textbf{\bibinfo{volume}{71}},
  \bibinfo{pages}{2650} (\bibinfo{year}{1993}).

\bibitem[{\citenamefont{H\"ochner}(2002)}]{hoechner}
\bibinfo{author}{\bibfnamefont{A.}~\bibnamefont{H\"ochner}}, Master's thesis,
  \bibinfo{school}{Z\"urich Universit\"at} (\bibinfo{year}{2002}).

\bibitem[{\citenamefont{Nandor et~al.}(1999)\citenamefont{Nandor, Martindale,
  Groves, Vyaselev, Pennington, Hults, and Smith}}]{nandor:99}
\bibinfo{author}{\bibfnamefont{V.~A.} \bibnamefont{Nandor}},
  \bibinfo{author}{\bibfnamefont{J.~A.} \bibnamefont{Martindale}},
  \bibinfo{author}{\bibfnamefont{R.~W.} \bibnamefont{Groves}},
  \bibinfo{author}{\bibfnamefont{O.~M.} \bibnamefont{Vyaselev}},
  \bibinfo{author}{\bibfnamefont{C.~H.} \bibnamefont{Pennington}},
  \bibinfo{author}{\bibfnamefont{L.}~\bibnamefont{Hults}}, \bibnamefont{and}
  \bibinfo{author}{\bibfnamefont{J.~L.} \bibnamefont{Smith}},
  \bibinfo{journal}{Phys. Rev. B} \textbf{\bibinfo{volume}{60}},
  \bibinfo{pages}{6907} (\bibinfo{year}{1999}).

\bibitem[{\citenamefont{Pines and Slichter}(1955)}]{ps55}
\bibinfo{author}{\bibfnamefont{D.}~\bibnamefont{Pines}} \bibnamefont{and}
  \bibinfo{author}{\bibfnamefont{C.~P.} \bibnamefont{Slichter}},
  \bibinfo{journal}{Phys. Rev.} \textbf{\bibinfo{volume}{100}},
  \bibinfo{pages}{1014} (\bibinfo{year}{1955}), \bibinfo{note}{which is known
  as the ``Wabash Cannonball Paper'' since it was written during the train ride
  to Detroit for the 1955 APS March Meeting (C. P. Slichter, private
  communication).}

\bibitem[{\citenamefont{Barrett et~al.}(1991)\citenamefont{Barrett, Martindale,
  Durand, Pennington, Slichter, Friedmann, Rice, and Ginsberg}}]{barret:91}
\bibinfo{author}{\bibfnamefont{S.~E.} \bibnamefont{Barrett}},
  \bibinfo{author}{\bibfnamefont{J.~A.} \bibnamefont{Martindale}},
  \bibinfo{author}{\bibfnamefont{D.~J.} \bibnamefont{Durand}},
  \bibinfo{author}{\bibfnamefont{C.~H.} \bibnamefont{Pennington}},
  \bibinfo{author}{\bibfnamefont{C.~P.} \bibnamefont{Slichter}},
  \bibinfo{author}{\bibfnamefont{T.~A.} \bibnamefont{Friedmann}},
  \bibinfo{author}{\bibfnamefont{J.~P.} \bibnamefont{Rice}}, \bibnamefont{and}
  \bibinfo{author}{\bibfnamefont{D.~M.} \bibnamefont{Ginsberg}},
  \bibinfo{journal}{Phys. Rev. Lett.} \textbf{\bibinfo{volume}{66}},
  \bibinfo{pages}{108} (\bibinfo{year}{1991}).

\bibitem[{\citenamefont{Pennington and Slichter}(1991)}]{penn:91}
\bibinfo{author}{\bibfnamefont{C.~H.} \bibnamefont{Pennington}}
  \bibnamefont{and} \bibinfo{author}{\bibfnamefont{C.~P.}
  \bibnamefont{Slichter}}, \bibinfo{journal}{Phys. Rev. Lett.}
  \textbf{\bibinfo{volume}{66}}, \bibinfo{pages}{381} (\bibinfo{year}{1991}).

\bibitem[{\citenamefont{Imai et~al.}(1992)\citenamefont{Imai, Slichter,
  Paulikas, and Veal}}]{imaisliT2G:92}
\bibinfo{author}{\bibfnamefont{T.}~\bibnamefont{Imai}},
  \bibinfo{author}{\bibfnamefont{C.~P.} \bibnamefont{Slichter}},
  \bibinfo{author}{\bibfnamefont{A.~P.} \bibnamefont{Paulikas}},
  \bibnamefont{and} \bibinfo{author}{\bibfnamefont{B.}~\bibnamefont{Veal}},
  \bibinfo{journal}{Appl. Magn. Reson.} \textbf{\bibinfo{volume}{3}},
  \bibinfo{pages}{729} (\bibinfo{year}{1992}).

\bibitem[{\citenamefont{Imai et~al.}(1993{\natexlab{a}})\citenamefont{Imai,
  Slichter, Paulikas, and Veal}}]{imaisliT2G:93}
\bibinfo{author}{\bibfnamefont{T.}~\bibnamefont{Imai}},
  \bibinfo{author}{\bibfnamefont{C.~P.} \bibnamefont{Slichter}},
  \bibinfo{author}{\bibfnamefont{A.~P.} \bibnamefont{Paulikas}},
  \bibnamefont{and} \bibinfo{author}{\bibfnamefont{B.}~\bibnamefont{Veal}},
  \bibinfo{journal}{Phys. Rev. B} \textbf{\bibinfo{volume}{47}},
  \bibinfo{pages}{R9158} (\bibinfo{year}{1993}{\natexlab{a}}).

\bibitem[{\citenamefont{Haase et~al.}(1999)\citenamefont{Haase, Morr, and
  Slichter}}]{haase:99}
\bibinfo{author}{\bibfnamefont{J.}~\bibnamefont{Haase}},
  \bibinfo{author}{\bibfnamefont{D.~K.} \bibnamefont{Morr}}, \bibnamefont{and}
  \bibinfo{author}{\bibfnamefont{C.~P.} \bibnamefont{Slichter}},
  \bibinfo{journal}{Phys. Rev. B} \textbf{\bibinfo{volume}{59}},
  \bibinfo{pages}{7191} (\bibinfo{year}{1999}).

\bibitem[{\citenamefont{Uldry et~al.}()\citenamefont{Uldry, Mali, Roos, and
  Meier}}]{uldryAni}
\bibinfo{author}{\bibfnamefont{A.}~\bibnamefont{Uldry}},
  \bibinfo{author}{\bibfnamefont{M.}~\bibnamefont{Mali}},
  \bibinfo{author}{\bibfnamefont{J.}~\bibnamefont{Roos}}, \bibnamefont{and}
  \bibinfo{author}{\bibfnamefont{P.~F.} \bibnamefont{Meier}},
  \emph{\bibinfo{title}{Anisotropy of the antiferromagnetic spin correlations
  in the superconducting state of yba$_2$cu$_3$o$_{7}$ and
  yba$_2$cu$_4$o$_{8}$}}, \bibinfo{howpublished}{cond-mat/0506245}.

\bibitem[{\citenamefont{Raffa et~al.}(1998)\citenamefont{Raffa, Ohno, Mali,
  Roos, Brinkmann, Conder, and Eremin}}]{raffa}
\bibinfo{author}{\bibfnamefont{F.}~\bibnamefont{Raffa}},
  \bibinfo{author}{\bibfnamefont{T.}~\bibnamefont{Ohno}},
  \bibinfo{author}{\bibfnamefont{M.}~\bibnamefont{Mali}},
  \bibinfo{author}{\bibfnamefont{J.}~\bibnamefont{Roos}},
  \bibinfo{author}{\bibfnamefont{D.}~\bibnamefont{Brinkmann}},
  \bibinfo{author}{\bibfnamefont{K.}~\bibnamefont{Conder}}, \bibnamefont{and}
  \bibinfo{author}{\bibfnamefont{M.}~\bibnamefont{Eremin}},
  \bibinfo{journal}{Phys. Rev. Lett.} \textbf{\bibinfo{volume}{81}},
  \bibinfo{pages}{5912} (\bibinfo{year}{1998}).

\bibitem[{\citenamefont{Curro et~al.}(1997)\citenamefont{Curro, Imai, Slichter,
  and Dabrowski}}]{curro}
\bibinfo{author}{\bibfnamefont{N.~J.} \bibnamefont{Curro}},
  \bibinfo{author}{\bibfnamefont{T.}~\bibnamefont{Imai}},
  \bibinfo{author}{\bibfnamefont{C.~P.} \bibnamefont{Slichter}},
  \bibnamefont{and}
  \bibinfo{author}{\bibfnamefont{B.}~\bibnamefont{Dabrowski}},
  \bibinfo{journal}{Phys. Rev. B} \textbf{\bibinfo{volume}{56}},
  \bibinfo{pages}{877} (\bibinfo{year}{1997}).

\bibitem[{\citenamefont{Tomeno et~al.}(1994)\citenamefont{Tomeno, Machi, Tai,
  Koshizuka, Kambe, Hayashi, Ueda, and Yasuoka}}]{tomeno:94}
\bibinfo{author}{\bibfnamefont{I.}~\bibnamefont{Tomeno}},
  \bibinfo{author}{\bibfnamefont{T.}~\bibnamefont{Machi}},
  \bibinfo{author}{\bibfnamefont{K.}~\bibnamefont{Tai}},
  \bibinfo{author}{\bibfnamefont{N.}~\bibnamefont{Koshizuka}},
  \bibinfo{author}{\bibfnamefont{S.}~\bibnamefont{Kambe}},
  \bibinfo{author}{\bibfnamefont{A.}~\bibnamefont{Hayashi}},
  \bibinfo{author}{\bibfnamefont{Y.}~\bibnamefont{Ueda}}, \bibnamefont{and}
  \bibinfo{author}{\bibfnamefont{H.}~\bibnamefont{Yasuoka}},
  \bibinfo{journal}{Phys. Rev. B} \textbf{\bibinfo{volume}{49}},
  \bibinfo{pages}{15327} (\bibinfo{year}{1994}).

\bibitem[{\citenamefont{Haase et~al.}(2002)\citenamefont{Haase, Slichter, and
  Milling}}]{haase:02}
\bibinfo{author}{\bibfnamefont{J.}~\bibnamefont{Haase}},
  \bibinfo{author}{\bibfnamefont{C.~P.} \bibnamefont{Slichter}},
  \bibnamefont{and} \bibinfo{author}{\bibfnamefont{C.~T.}
  \bibnamefont{Milling}}, \bibinfo{journal}{J. Supercond.}
  \textbf{\bibinfo{volume}{15}}, \bibinfo{pages}{339} (\bibinfo{year}{2002}).

\bibitem[{\citenamefont{Ohsugi et~al.}(1994)\citenamefont{Ohsugi, Y.Kitaoka,
  K.Ishida, G.Q.Zheng, and K.Asayama}}]{ohsugi}
\bibinfo{author}{\bibfnamefont{S.}~\bibnamefont{Ohsugi}},
  \bibinfo{author}{\bibnamefont{Y.Kitaoka}},
  \bibinfo{author}{\bibnamefont{K.Ishida}},
  \bibinfo{author}{\bibnamefont{G.Q.Zheng}}, \bibnamefont{and}
  \bibinfo{author}{\bibnamefont{K.Asayama}}, \bibinfo{journal}{J. Phys. Soc.
  Jpn} \textbf{\bibinfo{volume}{63}}, \bibinfo{pages}{700}
  (\bibinfo{year}{1994}).

\bibitem[{\citenamefont{Imai et~al.}(1993{\natexlab{b}})\citenamefont{Imai,
  Slichter, Yoshimura, and Kosuge}}]{imaisli:93}
\bibinfo{author}{\bibfnamefont{T.}~\bibnamefont{Imai}},
  \bibinfo{author}{\bibfnamefont{C.~P.} \bibnamefont{Slichter}},
  \bibinfo{author}{\bibfnamefont{K.}~\bibnamefont{Yoshimura}},
  \bibnamefont{and} \bibinfo{author}{\bibfnamefont{K.}~\bibnamefont{Kosuge}},
  \bibinfo{journal}{Phys. Rev. Lett.} \textbf{\bibinfo{volume}{70}},
  \bibinfo{pages}{1002} (\bibinfo{year}{1993}{\natexlab{b}}).

\bibitem[{\citenamefont{Moriya}(1956)}]{moriya:56}
\bibinfo{author}{\bibfnamefont{T.}~\bibnamefont{Moriya}},
  \bibinfo{journal}{Prog. Theor. Phys.} \textbf{\bibinfo{volume}{16}},
  \bibinfo{pages}{641} (\bibinfo{year}{1956}).

\bibitem[{\citenamefont{Haase et~al.}(2000)\citenamefont{Haase, Slichter,
  Stern, Milling, and Hinks}}]{haase:00}
\bibinfo{author}{\bibfnamefont{J.}~\bibnamefont{Haase}},
  \bibinfo{author}{\bibfnamefont{C.~P.} \bibnamefont{Slichter}},
  \bibinfo{author}{\bibfnamefont{R.}~\bibnamefont{Stern}},
  \bibinfo{author}{\bibfnamefont{C.~T.} \bibnamefont{Milling}},
  \bibnamefont{and} \bibinfo{author}{\bibfnamefont{D.~G.} \bibnamefont{Hinks}},
  \bibinfo{journal}{J. Supercond.} \textbf{\bibinfo{volume}{13}},
  \bibinfo{pages}{723} (\bibinfo{year}{2000}).

\bibitem[{\citenamefont{Curro et~al.}(2004)\citenamefont{Curro, Young,
  Schmalian, and Pines}}]{curropines}
\bibinfo{author}{\bibfnamefont{N.~J.} \bibnamefont{Curro}},
  \bibinfo{author}{\bibfnamefont{B.-L.} \bibnamefont{Young}},
  \bibinfo{author}{\bibfnamefont{J.}~\bibnamefont{Schmalian}},
  \bibnamefont{and} \bibinfo{author}{\bibfnamefont{D.}~\bibnamefont{Pines}},
  \bibinfo{journal}{Phys. Rev. B} \textbf{\bibinfo{volume}{70}},
  \bibinfo{pages}{235117} (\bibinfo{year}{2004}).

\bibitem[{\citenamefont{Imai et~al.}(1995)\citenamefont{Imai, Slichter, Cobb,
  and Markert}}]{imaisli:edoped}
\bibinfo{author}{\bibfnamefont{T.}~\bibnamefont{Imai}},
  \bibinfo{author}{\bibfnamefont{C.~P.} \bibnamefont{Slichter}},
  \bibinfo{author}{\bibfnamefont{J.~L.} \bibnamefont{Cobb}}, \bibnamefont{and}
  \bibinfo{author}{\bibfnamefont{J.~T.} \bibnamefont{Markert}},
  \bibinfo{journal}{J. Phys. Chem. Solids} \textbf{\bibinfo{volume}{56}},
  \bibinfo{pages}{1921} (\bibinfo{year}{1995}).

\bibitem[{\citenamefont{Harshman and Mills}(1992)}]{harsh}
\bibinfo{author}{\bibfnamefont{D.~R.} \bibnamefont{Harshman}} \bibnamefont{and}
  \bibinfo{author}{\bibfnamefont{A.~P.} \bibnamefont{Mills}},
  \bibinfo{journal}{Phys. Rev. B} \textbf{\bibinfo{volume}{45}},
  \bibinfo{pages}{10684} (\bibinfo{year}{1992}).

\end{thebibliography}
\end{document}